\documentclass[twocolumn,twocolumn,showpacs,aps,prstab,amsmath,amssymb]{revtex4-1}
\usepackage[latin9]{inputenc}
\usepackage{wrapfig}
\usepackage{textcomp}
\usepackage{amsmath}
\usepackage{graphicx}
\usepackage{esint}

\makeatletter

\providecommand{\tabularnewline}{\\}

\usepackage{graphicx}
\usepackage{bm}
\usepackage{booktabs}
\usepackage[usenames]{color}
\usepackage{hyphenat}
\usepackage{titlesec}

\renewcommand{\[}{\begin{equation}}
\renewcommand{\]}{\end{equation}}
\usepackage{wrapfig}
\renewcommand\thesection{\Roman{section}}

\makeatother

\begin{document}
\global\long\def\thesection{\arabic{section}}
 \numberwithin{equation}{section}

%\marginpar{BNL-112480-2016}

\setlength{\belowdisplayskip}{1pt} \setlength{\belowdisplayshortskip}{1pt} \setlength{\abovedisplayskip}{1pt} \setlength{\abovedisplayshortskip}{1pt}
\titlespacing*{\section} {0pt}{1mm plus 1mm minus .2mm}{1mm plus .2mm}
\titlespacing*{\subsection} {0pt}{1mm plus 1mm minus .2mm}{1mm plus .2mm}
\setlength{\textfloatsep}{1\baselineskip plus 0.2\baselineskip minus 0.5\baselineskip} 
\setlength{\intextsep}{1\baselineskip plus 0.2\baselineskip minus 0.5\baselineskip}

\title{A new method to probe the boundary where KAM tori persist by square
matrix}

\author{Li Hua Yu}

\affiliation{Brookhaven National Laboratory, Upton, NY 11973}
\begin{abstract}
The nonlinear dynamics of a system can be analyzed using a square
matrix. If off resonance, the lead vector of a Jordan chain in a left
eigenspace of the square matrix is an accurate action-angle variable
for sufficiently high power order. The deviation from constancy of
the action-angle variable provides a measure of the stability of a
trajectory. However, near resonance or the stability boundary, the
fluctuation increases rapidly and the lead vector no longer represents
an accurate action-angle variable. In this paper we show that near
resonance or stability boundary, it is possible to find a set of linear
combinations of the vectors in the degenerate Jordan chains as the
action-angle variables by an iteration procedure so that the fluctuation
is minimized. Using the Henon-Heiles problem as an example on resonance,
we show that when compared with conventional canonical perturbation
theory, the iteration leads to result in much more close agreement
with the forward integration, and the iteration is convergent even
very close to the stability boundary. It is further shown that the
action-angle variables found in the iteration can be used to find
another action-angle variables even much more closer to rigid rotations
(KAM invariants). The fast convergent result is not in the form of
polynomials, it is an exponential function with a rational function
in the exponent more similar to a Laurent series than a Taylor series.
Hence the method provides a new way to probe the boundary of the region
where KAM tori exist.
\end{abstract}


\maketitle
The field of nonlinear dynamics has very wide area of application
in science\cite{lieberman}. In general, the more common approach
is the forward numerical integration. To gain understanding, however,
one prefers an approximate analytical solution to extract relevant
information. ``Integrable systems in their phase space contain lots
of invariant tori and KAM Theory establishes persistence of such tori,
which carry quasi-periodic motions'' in small perturbations\cite{broer,porshel,arnold,wayne}.
But the theory does not provide the stability boundary. An important
issue is to find approximation to such KAM tori wherever they exist.
Among the many approaches to this issue we may mention canonical perturbation
theory, Lie algebra, power series, normal form\cite{lieberman,Gustavson,chirikov,dragt,berz,chao,forestnorm,brown},
etc. The results are often expressed as polynomials. However, for
increased perturbation, near resonance, or for large amplitude, the
solution of these perturbative approaches often lost precision.

The square matrix analysis developed recently \cite{nash,yu} has
a potential in exploring these area. In the following sections, we
first introduce the nonlinear dynamics square matrix equation, using
the Henon-Heiles problem\cite{Gustavson,henon} as an example. Then,
in Section 2, we show that when off resonance the lead vector of the
left Jordan chain is an approximate action-angle variable, and the
expression derived for the frequency shift is a rational function
rather than a polynomial. But on resonance, there are degenerate Jordan
chains, the action-angle variables are no longer the lead vectors
in the Jordan chains, they become the linear combinations of these
vectors. In Section 2 we show for small amplitude a differential equation
approach leads to an approximate solution, i.e., the approximate action-angle
variables. Expressed in these variables the exact solution nearly
represents rigid rotations but has amplitude and phase fluctuation.
As the amplitude increases the fluctuation increases, and the approximation
lost precision. 

In Section 3, we show that for a given approximate solution, we can
apply KAM theorem and use Fourier transform to find the coefficients
of the linear combinations to minimize the fluctuation. In Section
4, we show that when the fluctuation is small, we can write the equations
of motion in an exact form in terms of the action-angle variables
and treat the fluctuation as a perturbation to the rigid rotations
which serves as the zeroth order approximation. Then, in Section 5
we apply Fourier transform to solve this perturbation problem, and
find the first order approximation. In Section 6, we show that the
steps developed in section 3-5 form an iteration procedure. In each
iteration step, we solve a set of linear equations to improve the
the precision. Then, we use the Henon-Heiles problem as a numerical
example to compare the iteration result with the canonical perturbation
approach \cite{Gustavson}. Finally, in Section 7, we show that the
action-angle variables obtained by the iteration procedure, already
in very close agreement with forward integration, can be used immediately
to find another set of action-angle variables even much more closer
to rigid rotations. 

We summarize the result in the Conclusion: the numerical study shows
that when the iteration procedure is convergent, it leads to accurate
solution for a rather general type of nonlinear dynamics problem,
the convergence region provides information about the boundary where
KAM tori persist, and, since the perturbation is determined by the
ratio of fluctuation over the amplitude rather than amplitude itself,
it is different from, or might be even beyond the conventional canonical
perturbation theory. 

Our goal is to find the boundary where the KAM tori persist. The numerical
study suggests the border of the convergence region of the iteration
is very close to this boundary. However, our knowledge about the convergence
so far is limited to numerical study of the Henon-Heiles problem.
The relation between the convergence and the chaotic boundary is still
unknown, and an analytical analysis of this relation would be a very
important open issue.

\section{Introduction: square matrix equation for nonlinear dynamics}

We consider the equations of motion of a nonlinear dynamic system,
it can be expressed by a square matrix. We use the Henon-Heiles problem
\cite{Gustavson,henon} as an example, the Hamiltonian and the equations
of motion are 
\begin{align}
 & H=\frac{1}{2}(x^{2}+p_{x}^{2}+y^{2}+p_{y}^{2})+x^{2}y-\frac{1}{3}y^{3}\nonumber \\
 & \dot{x}=p_{x}\nonumber \\
 & \dot{y}=p_{y}\label{eq:I1}\\
 & \dot{p}_{x}=-x-2xy\nonumber \\
 & \dot{p}_{y}=-y-x^{2}+y^{2}\nonumber 
\end{align}
If we use the complex variables $z_{x}\equiv x-ip_{x},\,z_{x}^{*}\equiv x+ip_{x},z_{y}\equiv y-ip_{y},z_{y}^{*}\equiv y+ip_{y}$
to form a row of monomials $Z^{T}$$\equiv$ $\{z_{x},z_{x}^{*}$,
$z_{y},z_{y}^{*},$ $z_{x}^{2},z_{x}z_{x}^{*},z_{x}z_{y}$, $z_{x}z_{y}^{*},z_{x}^{*2}$,
$z_{x}^{*}z_{y}$,$z_{x}^{*}z_{y}^{*}$, $z_{y}^{2}$,$z_{y}z_{y}^{*}$,
$z_{y}^{*2}$,$z_{x}^{3}$, $z_{x}^{2}z_{x}^{*}$, $...$, $z_{y}^{*3}\}$,
we can write the following equations for the derivatives of the column
$Z$

{\small

\begin{align}
 & \dot{z}_{x}=iz_{x}+\frac{i}{2}z_{x}z_{y}+\frac{i}{2}z_{x}^{*}z_{y}+\frac{i}{2}z_{x}z_{y}^{*}+\frac{i}{2}z_{x}^{*}z_{y}^{*}\nonumber \\
 & \dot{z}_{x}^{*}=-iz_{x}^{*}-\frac{i}{2}z_{x}z_{y}-\frac{i}{2}z_{x}^{*}z_{y}-\frac{i}{2}z_{x}z_{y}^{*}-\frac{i}{2}z_{x}^{*}z_{y}^{*}\nonumber \\
 & \dot{z}_{y}=iz_{y}+\frac{i}{4}z_{x}^{2}+\frac{i}{2}z_{x}z_{x}^{*}+\frac{i}{4}z_{x}^{*2}-\frac{i}{4}z_{y}^{2}-\frac{i}{2}z_{y}z_{y}^{*}-\frac{i}{4}z_{y}^{*2}\nonumber \\
 & \dot{z}_{y}^{*}=-iz_{y}^{*}-\frac{i}{4}z_{x}^{2}-\frac{i}{2}z_{x}z_{x}^{*}-\frac{i}{4}z_{x}^{*2}+\frac{i}{4}z_{y}^{2}+\frac{i}{2}z_{y}z_{y}^{*}+\frac{i}{4}z_{y}^{*2}\label{eq:I2}\\
 & \frac{d}{dt}z_{x}^{2}=2z_{x}\dot{z}_{x}=2iz_{x}^{2}+iz_{x}^{2}z_{y}+iz_{x}z_{x}^{*}z_{y}+iz_{x}^{2}z_{y}^{*}+iz_{x}z_{x}^{*}z_{y}^{*}\nonumber \\
 & \frac{d}{dt}(z_{x}z_{x}^{*})=\frac{i}{2}z_{x}^{*2}z_{y}-\frac{i}{2}z_{x}^{2}z_{y}+\frac{i}{2}z_{x}^{*2}z_{y}^{*}-\frac{i}{2}z_{x}^{2}z_{y}^{*}\nonumber \\
 & ....\nonumber \\
 & \frac{d}{dt}z_{y}^{*3}=3z_{y}^{*2}\dot{z}_{y}^{*}=-3iz_{y}^{*3}\nonumber 
\end{align}

}As an example, we only keep the monomials to a power order of $n_{s}=3$.
The monomials are ordered according to power order from low to high.
Within each power order $n_{k}$, arrange the monomials so the first
variable power starts from power $n_{m}=n_{k}$ and followed by decreasing
power. For those terms with the first factor as monomial $z_{x}^{n_{m}}$,
other variables are arranged according to the same rule for a sequence
with power order $n_{k}-n_{m}$. One can show that the number of terms
of power order $n_{k}$ is $(n_{k}+1)(n_{k}+2)(n_{k}+3)/6$, so for
power $n_{k}=1,2,3$ the numbers are 4,10,20 respectively. The total
number of terms up to power $n_{s}$ is $4+10+20=(n_{s}+1)(n_{s}+2)(n_{s}+3)(n_{s}+4)/24-1=34$.
Then the differential equations Eq.(\ref{eq:I1}) can be represented
approximately by a large $34\times34$ square matrix $M$ as $\dot{{Z}}=MZ.$

The square matrix $M$ is upper-triangular, and has the following
form with the dimension of its sub-matrices determined by the number
of terms for each power order 4,10,20: 
\[
M=\begin{bmatrix}M_{11} & M_{12} & M_{13}\\
0 & M_{22} & M_{23}\\
0 & 0 & M_{33}
\end{bmatrix}
\]
The diagonal blocks $M_{11}$, $M_{22}$, $M_{33}$ are all diagonal
matrices with dimension $4\times4$, $10\times10$, $20\times20$
respectively, thus $M$ is a $34\times34$ upper-triangular matrix.
The diagonal elements of $M_{11}$, $M_{22}$, $M_{33}$ are $\{i,-i,i,-i\}$,
$\{2i,0,2i,0,-2i,0,-2i,2i,0,-2i\}$,$\{3i,i,3i,...,-3i\}$, respectively.
For the sake of space, we only give a part of the list. But one can
recognize the pattern for the diagonal elements. If we only keep the
linear terms in the first 4 rows of Eq.(\ref{eq:I2}), and denote
the eigenvalues of the linear part of the equations for $z_{x}$ and
$z_{y}$ as $i\mu_{x}$ and $i\mu_{y}$ respectively, then $\mu_{x}=\mu_{y}\equiv\mu=1$,
the system is on resonance. The solution would be $z_{x}=z_{y}=e^{i\mu t},z_{x}^{*}=z_{y}^{*}=e^{-i\mu t}$.
This is the small amplitude limit of the solution. If we substitute
these into $Z$, then $\dot{Z}(t=0)$ gives the diagonal elements.

The dimension of the off-diagonal blocks $M_{12}$, $M_{13}$, $M_{23}$
are $4\times10$, $4\times20$, $10\times20$ respectively. Since
there are no third power terms in the first 4 rows of Eq.(\ref{eq:I2})
, $M_{13}=0$.{\small 
\setlength\arraycolsep{1.5pt}

\begin{equation}
\begin{split} & M_{12}=\begin{bmatrix}0 & 0 & \frac{i}{2} & \frac{i}{2} & 0 & \frac{i}{2} & \frac{i}{2} & 0 & 0 & 0\\
0 & 0 & -\frac{i}{2} & -\frac{i}{2} & 0 & -\frac{i}{2} & -\frac{i}{2} & 0 & 0 & 0\\
\frac{i}{4} & \frac{i}{2} & 0 & 0 & \frac{i}{4} & 0 & 0 & -\frac{i}{4} & -\frac{i}{2} & -\frac{i}{4}\\
-\frac{i}{4} & -\frac{i}{2} & 0 & 0 & -\frac{i}{4} & 0 & 0 & \frac{i}{4} & \frac{i}{2} & \frac{i}{4}
\end{bmatrix},\ M_{23}=\\
 & \begin{bmatrix}0 & 0 & i & i & 0 & i & i & 0 & 0 & 0 & 0 & 0 & 0 & 0 & 0 & 0 & 0 & 0 & 0 & 0\\
0 & 0 & -\frac{i}{2} & -\frac{i}{2} & 0 & 0 & 0 & 0 & 0 & 0 & 0 & \frac{i}{2} & \frac{i}{2} & 0 & 0 & 0 & 0 & 0 & 0 & 0\\
... & ... & ... & ... & ... & ... & ... & ... & ... & ... & ... & ... & ... & ... & ... & ... & ... & ... & ... & ...\\
0 & 0 & 0 & -\frac{i}{2} & 0 & 0 & -i & 0 & 0 & 0 & 0 & 0 & -\frac{i}{2} & 0 & 0 & 0 & 0 & \frac{i}{2} & i & \frac{i}{2}
\end{bmatrix}
\end{split}
\label{eq:(1.5)-1}
\end{equation}

}The dimension of $M$ increases rapidly with the power order $n_{s}.$
However, because $M$ is upper-triangular, it is straight forward
to find its Jordan subspaces with much lower dimensions, and the eigenvalues
of these eigenspaces are the diagonal elements. For example, for $n_{s}=5$,
$M$ is a $125\times125$ matrix. The lengths of the Jordan chains
are 3, 2, 1\cite{yu}. Each chain forms the basis of an invariant
subspace. Only for the longest chains with eigenvalue $i\mu_{x}$
or $i\mu_{y}$, the lead vector has all the monomials from linear
terms to power $n_{s}$. For all other chains, the lead vectors have
only higher power terms. Hence our study is focused on these longest
chains. In general, the Jordan chains are not uniquely defined. However
there is a way to define them uniquely so the lead vector for each
chain with those terms of form $z_{x}(z_{x}z_{x}^{\ast})^{m_{x}}(z_{y}z_{y}^{\ast})^{m_{y}}$,
or $z_{y}(z_{x}z_{x}^{\ast})^{m_{x}}(z_{y}z_{y}^{\ast})^{m_{y}}$,
i.e., the terms of form of $z_{x}$,$z_{y}$ times an invariant monomial
removed(see in Appendix E of \cite{yu}). For Henon-Heiles problem,
the system is on resonance with $\mu_{x}=\mu_{y}=1$, so if $n_{s}=5$,
the two invariant subspace formed by the two longest chains with the
eigenvalue $i\mu_{x}$ and $i\mu_{y}$ joined into one invariant subspace
of dimension 6, much lower than the dimension of $M$.

\section{\label{sec:I}The Action Angle Variables based on Jordan Decomposition
of Square matrix}

For a square matrix equation $\dot{Z}=MZ$ as discussed in \cite{yu},
for one Jordan chain $U$ in the left eigenspace of $M$, we have
\begin{equation}
\begin{split} & UM=NU\\
 & \frac{d(UZ)}{dt}=U\dot{Z}=UMZ=NUZ
\end{split}
\label{eq:(1.1)}
\end{equation}
where $U$ is a rectangular matrix with each row a left eigenvector
in the Jordan chain, arranged in the order of the chain so the first
row is its lead vector. $N\equiv i\mu+\tau$ is the Jordan form in
one of the Jordan blocks we are interested in, i.e., the chain of
generalized eigenvectors of $M$ with eigenvalue $i\mu$. Now introduce
$W\equiv UZ$, with 
\begin{equation}
W=\begin{bmatrix}\,w_{0}\\
\,w_{1}\\
\,w_{2}\\
...
\end{bmatrix}\text{{\ and\ }}\tau W=\begin{bmatrix}0 & 1 & 0 & ... & 0\\
0 & 0 & 1 & ... & 0\\
0 & 0 & ... & ... & 0\\
0 & 0 & 0 & ... & 1\\
0 & 0 & 0 & ... & 0
\end{bmatrix}\begin{bmatrix}\,w_{0}\\
\,w_{1}\\
\,w_{2}\\
...
\end{bmatrix}=\begin{bmatrix}\,w_{1}\\
\,w_{2}\\
...\\
\,0
\end{bmatrix}\label{eq:(1.2)}
\end{equation}
where $w_{0}=u_{0}Z,\,w_{1}=u_{1}Z,\ldots$ are the projection of
vector $Z$ onto the eigenvectors $u_{0},\,u_{1},\ldots$(the rows
of $U$). During the motion, $Z$ rotates in the space of all polynomials
within a given power order $n_{s}$. Hence the vectors $w_{0},\,w_{1},\,\ldots$
also rotate in the subspace with eigenvalue $i\mu$. From Eq.(\ref{eq:(1.1)}),
we have 
\begin{equation}
\begin{split} & \dot{W}=(i\mu+\tau)W\end{split}
\label{eq:(1.3)}
\end{equation}
When far away from resonances, $W$ is approximately an eigenvector
of $\tau$. That is, it is a ``coherent'' state \cite{glauber,sudashan}
with eigenvalue $i\phi$, as discussed in \cite{yu}. $\phi$ is nearly
a constant representing the amplitude dependent frequency shift from
the zero amplitude frequency $\mu$. For this case, in Eq.(\ref{eq:(1.3)})
$\tau$ is approximately replaced by $i\phi$: $\dot{w}_{0}\approx i(\mu+\phi)w_{0},\dot{w}_{1}$
$\approx i(\mu+\phi)w_{1},\cdots$, with $\phi\approx w_{1}/w_{0}\approx w_{1}/w_{2}\approx\cdots$.
Hence the frequency shift $\phi$ is a rational function rather than
a polynomial. Each row of $W$ in Eq.(\ref{eq:(1.2)}) is an approximation
of the action-angle variable. The lead vector $w_{0}$ has terms from
linear to high power terms, $w_{1}$ has only terms of power higher
than 2, while all other $w_{j}$ with $j>1$ has only terms of power
higher than 4. Hence we can write $v\equiv re^{i\theta}\approx w_{0}$
as the action-angle variable approximation while all other $w_{j}$s
are less acurate action-angle variables with only higher power terms.

For the case of two variables $x$ and $y$, we consider two eigenspaces
and find the Jordan chain $W_{x}$ $\{w_{x0}=u_{x0}Z,$ $w_{x1}=u_{x1}Z,\ldots$\}
, and Jordan chain $W_{y}$ \{$w_{y0}=u_{y0}Z,\,w_{y1}=u_{y1}Z,\ldots$
\}, with eigenvalues $i\mu_{x,}i\mu_{y}$ respectively. Similarly
we find two independent action-angle variable approximations $v_{x}\equiv r_{x}e^{i\theta_{x}}\approx w_{x0}$,
$v_{y}\equiv r_{y}e^{i\theta_{y}}\approx w_{y0}$. The rotations of
these two vectors form two independent rigid rotations with approximately
constant phase advance rates $\dot{\theta}_{x}=\mu_{x}+\phi_{x}$
and $\dot{\theta}_{y}=\mu_{y}+\phi_{y}$ respectively. As described
in \cite{yu}, these variables provide excellent solution to the nonlinear
dynamics problem when off resonances.

However, when on resonance, $\mu_{x}=\mu_{y}=\mu$, the two blocks
$W_{x}$ and $W_{y}$ are degenerate. They are no longer approximate
eigenvectors of $\tau$. In spite of this, because the projection
of the trajectory $Z$ onto the eigenspace $\mu$ must remain in this
invariant subspace, it must be a linear combination of $w_{x0}$ ,$w_{y0}$,
$w_{x1}$, and $w_{y1}\ldots$. In addition, KAM theory states that
under small perturbation there must be a region around the fixed point
where the invariant tori are stable, hence the corresponding linear
combinations must have well defined frequencies, or, form a coherent
state.

In the following, when we refer to the linear combinations, it is
equivalent that we refer to the approximate action-angle variables.
Thus the main issue in finding the solution becomes finding the coefficients
of such linear combinations for which the corresponding action-angle
variables evolve in a way that the deviation (i.e., the fluctuation)
from rigid rotations is minimized. Thus, for each pair of approximate
action-angle variables we associate it with an approximate trajectory,
and a pair of rigid rotations. The deviation of the trajectory from
the rigid rotations is to be minimized.

For the case on resonance in the Henon-Heiles problem of two variables
$x$ and $y$, we are searching for two sets of linear combinations
for two approximate action-angle variables as the solution. In the
search for the linear combinations of the vectors in the eigenspace,
we first consider the case of small amplitude. We neglect $w_{x1},w_{y1}\ldots$
, which only have high power terms. The vectors $w_{x1},w_{y1}\ldots$
have no linear terms, their contribution becomes important only for
motion with large amplitude, as we shall describe later. Thus we look
for coefficients in the linear combination $v\equiv a_{1}w_{x0}+a_{2}w_{y0}\text{ }$
such that $v=re^{i\omega t}$, where $r$ and $\omega$ are constant.
Thus $v$ , as an approximate action-angle variable (in the following,
to be brief, we abbreviate it as ``action''), satisfies the equations
\begin{equation}
\begin{split} & \dot{v}=i\omega v\text{ }\\
 & \ddot{{v}}=i\omega\dot{{v}}
\end{split}
\text{ }\hspace{1cm}\label{eq:(1.4)}
\end{equation}
To derive an equation for $a_{1},a_{2}$, we substitute $v\equiv a_{1}w_{x0}+a_{2}w_{y0}\text{ }$
into Eq.(\ref{eq:(1.4)}). To calculate the derivatives, write Eq.(\ref{eq:(1.3)})
in an explicit matrix form, using the property of the Jordan matrix
$\tau$, we found $\dot{w}_{0}=i\mu\,w_{0}+w_{1}$and $\ddot{w}_{0}=(i\mu)^{2}w_{0}+2i\mu\,w_{1}+w_{2}$.
Applying these to the two degenerate chains $W_{x}$ and $W_{y}$
respectively, we get \vspace{-0.05em}
\begin{align}
 & \dot{v}=(i\mu\,w_{x0}+w_{x1})a_{1}+(i\mu\,w_{y0}+w_{y1})a_{2}\nonumber \\
 & \ddot{v}=((i\mu)^{2}w_{x0}+2i\mu\,w_{x1}+w_{x2})a_{1}\label{eq:1.9}\\
 & \qquad+((i\mu)^{2}w_{y0}+2i\mu\,w_{y1}+w_{y2})a_{2}\nonumber 
\end{align}
Substitute each row of Eq.(\ref{eq:1.9}) into Eq.(\ref{eq:(1.4)}),
let $i\phi=i\omega-i\mu$, we obtain an eigenequation
\begin{equation}
\begin{split}\begin{bmatrix}w_{x0} & w_{y0}\\
w_{x1} & w_{y1}
\end{bmatrix}^{-1}\begin{bmatrix}w_{x1} & w_{y1}\\
w_{x2} & w_{y2}
\end{bmatrix}A=i\phi A\end{split}
\text{ }{\thinspace with\thinspace}\text{ }\hspace{1mm}A\equiv\begin{bmatrix}a_{1}\\
a_{2}
\end{bmatrix}\label{14}
\end{equation}
This is a generalization of the off-resonance frequency shift $i\phi=\frac{w_{1}}{w_{0}}$.
Now it is an eigenequation with two eigenvalues $\phi_{1}\equiv\omega_{1}-\mu$,
$\phi_{2}\equiv\omega_{2}-\mu$. $w_{x0}$,$w_{y0}$,$w_{x1}$,$w_{y1}$,$w_{x2}$,$w_{y2}$
are polynomials of $x$,$p_{x}$,$y$,$p_{y}$. For a set of initial
values $x_{0}$,$p_{x0}$,$y_{0}$,$p_{y0}$, Eq.(\ref{14}) has two
eigenvectors $A_{1},A_{2}$, which give the two sets of coefficients
$\{a_{11},a_{12}\}$,$\{a_{21},a_{22}\}$ in $v\equiv a_{1}w_{x0}+a_{2}w_{y0}\text{ }$,
corresponding to the two functions $v_{1}\equiv r_{1}e^{i\theta_{1}}$,$v_{2}\equiv r_{2}e^{i\theta_{2}}$.
For small amplitude, they are excellent first order approximation
of the action-angle variables. Thus $\theta_{1}\approx\omega_{1}t+\theta_{10}=(\mu+\phi_{1})t+\theta_{10}$
and $\theta_{2}\approx\omega_{2}t+\theta_{20}=(\mu+\phi_{2})t+\theta_{20}$
represent two independent rigid rotations, as required by KAM theorem.
For small amplitude, the linear combination coefficients $a_{1},a_{2}$
are determined from the initial condition, and the deviation from
the rigid rotations is negligibly small.

\section{\label{sec:CalLin}Calculation of Linear Combinations for Action-Angle
Variables using a known trajectory}

In the case of increased amplitude, the linear combination coefficients
can no longer be determined by the initial value $x_{0}$,$p_{x0}$,$y_{0}$,$p_{y0}$
as in Eq.(\ref{14}) but by $x$,$p_{x}$,$y$,$p_{y}$ in a larger
neighborhood near the fixed point. The high power terms in Eq.(\ref{eq:(1.1)}),
which are truncated in the construction of the square matrix in Eq.(\ref{eq:(1.3)}),
serve as a perturbation to the rigid rotations as described in Poincare-Birkhoff
theorem \cite{brown}. Thus the main issue in finding the solution
now becomes finding the coefficients of the linear combinations for
large amplitude in a perturbation theory .

In the following we shall first show that if we have a numerical forward
integration of the dynamic equations, i.e., if we have the trajectory,
we can apply KAM theory and use Fourier expansion to determine the
linear combinations that approximate rigid rotations. However, since
our goal is to find the solution without the forward integration,
later we shall not use the trajectory given by the forward integration.
Instead, we use an approximate trajectory to determine the linear
combinations approximately. The first approximate trajectory itself
can be obtained by the method outlined in the Section \ref{sec:I},
with the rigid rotation calculated from the linear combinations given
by Eq.(\ref{14}). The trajectory is going to be improved by an iteration
procedure given in later sections.

KAM theory states that \cite{broer,porshel} under small perturbation
there must be a domain (a Cantor set with positive measure) around
the fixed point where the invariant tori (the perturbed rotation)
are stable, and represent quasi-periodic motions. But the theory does
not tell where the region is extended to. Our goal is to probe the
boundary of this region. Within this boundary, according to Arnold's
theorem\cite{arnold}, there is a variable transformation from the
perturbed rotation to rigid rotation (see detailed explanation in
\cite{wayne}, where the perturbed rotation is referred to as being
conjugated to the rigid rotation by the variable transform). In our
notation, there exists a transformation from the trajectory $x,p_{x},y,p_{y}$
to the rigid rotation $v_{1}(x,p_{x},y,p_{y})=r_{1}e^{i\omega_{1}t},$
$v_{2}(x,p_{x},y,p_{y})=r_{2}e^{i\omega_{2}t}$, i.e., the action-angle
variables. As an inverse function, the coordinates $x$,$p_{x}$,$y$,$p_{y}$
are the functions of $\theta_{1}\equiv\omega_{1}t$, $\theta_{2}\equiv\omega_{2}t$
(modulo $2\pi$), hence the eigenvectors $w_{x0}$ ,$w_{y0}$, $w_{x1}$,
$w_{y1}\ldots$ can also be expanded in terms of $\theta_{1}$, $\theta_{2}$.
This property of a trajectory as a function of $t$ can be represented
by a periodic function of $\theta_{1}$, $\theta_{2}$ is critically
important, so in the following search for approximate solutions, we
limit them to periodic functions of $\theta_{1}$, $\theta_{2}$ (modulo
$2\pi$) only. Our goal is to not only find the solution near the
stability boundary, but more importantly, transform it into a specific
form of rigid rotation using this property.

For simplicity in writing, if we choose $n_{v}$ eigenvectors for
the linear combinations, we label them as $w_{j}$ with $j=1,2,..,n_{v}$.
For example, for Eq.(\ref{eq:(1.4)}), $n_{v}=2$, $w_{1}\equiv w_{x0},w_{2}\equiv w_{y0}$.
We have the expansion

\begin{equation}
w_{j}(\theta_{1},\theta_{2})=\sum_{n,m}\widetilde{w}_{jnm}e^{in\theta_{1}}e^{im\theta_{2}}\qquad(j=1,2,..n_{v})\label{eq:10}
\end{equation}
Now we look for linear combinations $a_{1j}$,$a_{2j}$ to construct
the two approximate action-angle variables $v_{1},v_{2}$
\begin{align}
v_{l} & =\sum_{j=1}^{n_{v}}a_{lj}w_{j}=\sum_{n,m}\left(\sum_{j=1}^{n_{v}}a_{lj}\widetilde{w}_{jnm}\right)e^{in\theta_{1}}e^{im\theta_{2}}\label{eq:11}\\
 & \equiv\sum_{n,m}\widetilde{v}_{lnm}e^{in\theta_{1}}e^{im\theta_{2}}\qquad(l=1,2)\nonumber 
\end{align}
The Fourier coefficient for spectral line $n\omega_{1}+m\omega_{2}$
is $\widetilde{v}_{lnm}=\sum_{j}a_{lj}\widetilde{w}_{jnm}$. We choose
$a_{lj}$ such that $\widetilde{v}_{110}=1$, $\widetilde{v}_{201}=1$,
and define $\widetilde{v}_{1nm}=\epsilon_{1nm}$, for all ${n,m}$
except ${n=1,m=0}$, and $\widetilde{v}_{2nm}=\epsilon_{2nm}$ for
all ${n,m}$ except for ${n=0,m=1}$. $\epsilon_{lnm}$ represents
fluctuation. Among all possible values for $a_{lj}$, the one with
minimized fluctuation most closely represents the rigid rotations.
In general, we have a minimization problem for a function $g_{0}$
quadratic in $a_{lj}$ with constraints $g_{1},g_{2}$:
\begin{equation}
\begin{aligned} & g_{0}(a_{lj})=\sum_{\tiny{\begin{matrix}n,m\\
|n-1|+|m|\neq0
\end{matrix}}}|\epsilon_{1nm}|^{2}+\sum_{\tiny{\begin{matrix}n,m\\
|n|+|m-1|\neq0
\end{matrix}}}|\epsilon_{2nm}|^{2}\\
 & g_{1}(a_{lj})=\widetilde{v}_{110}-1=0\\
 & g_{2}(a_{lj})=\widetilde{v}_{201}-1=0
\end{aligned}
\label{eq:12}
\end{equation}
If $g_{0}=0$, then $\epsilon_{lnm}$ are all zero, $v_{l}=e^{i\omega_{l}t}$
has a single frequency $\omega_{l}$, and $v_{1},v_{2}$ would be
exact rigid rotations, representing perfect KAM tori. KAM theory states
that there must be a neighborhood of the integrable solution where
the invariant tori persist, so there should be a solution for which
when the square matrix is not truncated, i.e., when $n_{s}$ and $n_{v}$
approach infinity, the fluctuation vanishes, and $g_{0}$ approaches
zero. However, in a real example, when the matrix is truncated at
certain order, the fluctuation would not vanish. Thus we assume the
initial condition is such that it is in the region where the KAM tori
persist. And, for a finite power order $n_{s}$ and eigenvector number
$n_{v}$, we minimize the fluctuation $g_{0}$ to approximate the
KAM tori. 

Use Lagrangian multiplier $\lambda_{1}$, $\lambda_{2}$, the minimization
problem is reduced to solving $2n_{v}+2$ linear equations for $2n_{v}+2$
unknown $a_{lj,}$$\lambda_{1},\lambda_{2}$:

{% \abovedisplayskip=0pt%
\begin{equation}
\begin{aligned} & \frac{\partial g_{0}}{\partial a_{lj}}+\lambda_{1}\frac{\partial g_{1}}{\partial a_{lj}}+\lambda_{2}\frac{\partial g_{2}}{\partial a_{lj}}=0\qquad(l=1,2;\ j=1,2,..n_{v})\\
 & g_{1}=0,g_{2}=0
\end{aligned}
\label{eq:13}
\end{equation}
}The solution of Eq.(\ref{eq:13}) is straight forward, and gives
the linear combinations $a_{1k},a_{2k}$
\begin{align}
 & a_{1k}=\frac{\sum_{j}\left(F_{1}^{-1}\right)_{kj}\widetilde{w}_{j10}^{*}}{\sum_{m,j}\widetilde{w}_{m10}\left(F_{1}^{-1}\right)_{mj}\widetilde{w}_{j10}^{*}}\qquad(m,j,k=1,2,..n_{v})\nonumber \\
 & a_{2k}=\frac{\sum_{j}\left(F_{2}^{-1}\right)_{kj}\widetilde{w}_{j01}^{*}}{\sum_{m,j}\widetilde{w}_{m01}\left(F_{2}^{-1}\right)_{mj}\widetilde{w}_{j01}^{*}}\qquad\text{{with}}\label{eq:15}\\
 & \text{\ensuremath{\left(F_{1}\right)_{jk}\equiv\sum_{\tiny{\begin{matrix}n,m\\
 |n-1|+|m|\neq0 
\end{matrix}}}\widetilde{w}_{jnm}^{*}\widetilde{w}_{knm}}}\nonumber \\
 & \text{\ensuremath{\left(F_{2}\right)_{jk}\equiv\sum_{\tiny{\begin{matrix}n,m\\
 |n|+|m-1|\neq0 
\end{matrix}}}\widetilde{w}_{jnm}^{*}\widetilde{w}_{knm}}\quad}\nonumber 
\end{align}
For a given approximate trajectory $x$,$p_{x}$,$y$,$p_{y}$ as
function of $\theta_{1}$, $\theta_{2}$, the linear combinations
$a_{1k},a_{2k}$ Eq.(\ref{eq:15}) determine the approximate action-angle
variables $v_{1},v_{2}$ with minimized fluctuation Eq.(\ref{eq:12}),
so they approximately represent rigid rotations. In the following
sections we shall develop a perturbation theory based on these approximate
action-angle variables to obtain more accurate solution, i.e, more
accurate trajectory.

\section{\label{sec:IIIPerturbation}Perturbation based on action-angle approximation }

To formulate the perturbation problem, in the action-angle variables
$v_{1}\equiv r_{1}e^{i\theta_{1}}$,$v_{2}\equiv r_{2}e^{i\theta_{2}}$
we keep $r_{1}$,$r_{2}$ to be positive constants but take the deviation
from the rigid rotations into account by the assumption that $\theta_{1}$,
$\theta_{2}$ as functions of time having not only linear terms proportional
to time, also perturbation terms with small phase fluctuation (the
real part of $\theta$) and amplitude fluctuation (imaginary part
of $\theta$). For simplicity of writing, we use only two left eigenvectors
($n_{v}=2$) in the following, as in Eq.(\ref{eq:(1.4)}). But in
later sections we shall use $n_{v}=4$ for more accurate solution.
We consider the relation between the 4 variables, $\theta_{1}$, $\theta_{2}$,
their complex conjugate $\theta_{1}^{*}$, $\theta_{2}^{*}$, and
$x$,$p_{x}$,$y$,$p_{y}$ as a variable transformation{\small
\begin{equation}
\begin{split}v_{1}(\theta_{1},\theta_{2})\equiv r_{1}e^{i\theta_{1}}\equiv a_{11}w_{x0}(x,p_{x},y,p_{y})+a_{12}w_{y0}(x,p_{x},y,p_{y})\\
v_{2}(\theta_{1},\theta_{2})\equiv r_{2}e^{i\theta_{2}}\equiv a_{21}w_{x0}(x,p_{x},y,p_{y})+a_{22}w_{y0}(x,p_{x},y,p_{y})
\end{split}
\label{eq:19}
\end{equation}
}Hence we consider the right hand side of this equation as an implicit
function of $\theta_{1}$, $\theta_{2}$. The time derivative of Eq.(\ref{eq:19})
gives
\begin{equation}
\begin{aligned} & i\dot{\theta}_{1}v_{1}=\dot{v}_{1}=a_{11}\dot{w}_{x0}+a_{12}\dot{w}_{y0}\\
 & i\dot{\theta}_{2}v_{2}=\dot{v}_{2}=a_{21}\dot{w}_{x0}+a_{22}\dot{w}_{y0}
\end{aligned}
\label{eq:20}
\end{equation}
We consider Eq.(\ref{eq:19}) as a definition, hence when the derivatives
in the right hand side of Eq.(\ref{eq:20}) are calculated exactly
it is an \textbf{exact equation of motion} for the dynamic variables
$\theta_{1}$, $\theta_{2}$ . If in the right hand side of Eq.(\ref{eq:20})
we use the property of the Jordan matrix $\tau$ to calculate the
derivatives as in Eq.(\ref{eq:1.9}), because the derivatives such
as $\dot{w}_{x0}=i\mu\,w_{x0}+\tau w_{x0}=$ $i\mu\,w_{x0}+w_{x1}=$
$i(\mu+\frac{w_{x1}}{w_{x0}})\,w_{x0}\equiv i(\mu+\phi_{x})\,w_{x0}$
are based on Jordan decomposition of the square matrix, while the
matrix is truncated by a given finite power order $n_{s}$, it would
be an approximation. However, we may use the chain rule for the derivatives.
Since $w_{x0},w_{y0}$ are polynomials, their partial derivatives
such as $\partial w_{x0}/\partial x,\cdots$ are polynomials, while
the derivatives $\dot{x}$, $\dot{p}_{x}$,$\dot{y}$,$\dot{p}_{y}$,
are given by the equation of motion Eq.(\ref{eq:I1}) as the functions
of $x$,$p_{x}$,$y$,$p_{y}$, the right hand side of Eq.(\ref{eq:20})
can be derived exactly as function of $x$,$p_{x}$,$y$,$p_{y}$,
which in turn is an implicit function of $\theta_{1}$, $\theta_{2}$.
Thus we can rewrite Eq.(\ref{eq:20}) as \textbf{exact equations}
\begin{equation}
\begin{alignedat}{1} & \dot{{\theta}_{1}}=\mu+\phi_{1}(\theta_{1},\theta_{2})=\mu+\bar{\phi}_{1}+(\phi_{1}-\bar{\phi}_{1})\equiv\omega_{1}+\Delta\phi_{1}\\
 & \dot{{\theta}_{2}}=\mu+\phi_{2}(\theta_{1},\theta_{2})=\mu+\bar{\phi}_{2}+(\phi_{2}-\bar{\phi}_{2})\equiv\omega_{2}+\Delta\phi_{2}\\
 & \phi_{1}(\theta_{1},\theta_{2})\equiv-i(\dot{w}_{x0}+a_{1}\dot{w}_{y0})v_{1}^{-1}-\mu\\
 & \phi_{2}(\theta_{1},\theta_{2})\equiv-i(\dot{w}_{x0}+a_{2}\dot{w}_{y0})v_{2}^{-1}-\mu
\end{alignedat}
\label{eq:21}
\end{equation}
where we define $\bar{\phi}_{1}$ as the average value of $\phi_{1}$
over long time. We have absorbed the contribution of the constant
term $\bar{\phi}_{1}$ in $\phi_{1}(t)$ into the steady phase advance
rate $\omega_{1}\equiv\mu+\bar{\phi}_{1}$ of $v_{1}$, while the
deviation from the rigid rotation is represented by $\Delta\phi_{1}\equiv\phi_{1}-\bar{\phi}_{1}$,
with its real part as the phase fluctuation, and its imaginary part
as the amplitude fluctuation. $\bar{\phi}_{2}$, $\Delta\phi_{2}$,
and $\omega_{2}$ are similarly defined and calculated.

With this provision, the two terms $\Delta\phi_{1}$,$\Delta\phi_{2}$
serve as a perturbation to two independent rigid rotations. When they
are neglected, we obtain the zeroth order approximation, i.e. the
rigid rotations:
\begin{equation}
\begin{split}v_{1}^{(0)}\equiv r_{1}e^{i\theta_{1}^{(0)}(t)}\hspace{1em}\text{with\,}\,\theta_{1}^{(0)}(t)\equiv\omega_{1}t+\theta_{10}\\
v_{2}^{(0)}\equiv r_{2}e^{i\theta_{2}^{(0)}(t)}\hspace{1em}\text{with }\theta_{2}^{(0)}(t)\equiv\omega_{2}t+\theta_{20}
\end{split}
\label{eq:22}
\end{equation}
where $r_{1},r_{2},\theta_{10},\theta_{20}$ are the initial amplitudes
and phases of the actions $v_{1},v_{2}$. Let $\Delta\theta_{1},\Delta\theta_{2}$
represent the fluctuation:
\begin{equation}
\begin{split}\theta_{1}\equiv\theta_{1}^{(0)}+\Delta\theta_{1}=\omega_{1}t+\theta_{10}+\Delta\theta_{1}\\
\theta_{2}\equiv\theta_{2}^{(0)}+\Delta\theta_{2}=\omega_{2}t+\theta_{20}+\Delta\theta_{2}
\end{split}
\label{eq:23}
\end{equation}
To calculate the fluctuation, we approximate $\Delta\theta_{1}(t)$
by its first order approximation $\Delta\theta_{1}^{(1)}(t)$ as follows.

From Eq.(\ref{eq:21}) we have
\begin{equation}
\begin{split} & \theta_{1}^{(1)}(t)\equiv\theta_{1}^{(0)}(t)+\Delta\theta_{1}^{(1)}(t)\\
 & \Delta\theta_{1}^{(1)}(t)=\int_{0}^{t}dt'\thinspace(\phi_{1}(\theta_{1}^{(0)}(t'),\theta_{2}^{(0)}(t'))-\bar{\phi}_{1})
\end{split}
\label{eq:24}
\end{equation}
That is, we replace the phases in the right hand side of Eq.(\ref{eq:21})
by its zero order approximation $\theta_{1}^{(0)}(t'),\theta_{2}^{(0)}(t')$.
Notice that under this approximation, as $t'$ approaches infinity
$\theta_{1}^{(0)}(t'),\theta_{2}^{(0)}(t')$ scan through the plane
of the real part of $\theta_{1}$, $\theta_{2}$ (modulo $2\pi$)
uniformly, so $\bar{\phi}_{1}$ is the average of $\phi_{1}$ over
the plane of the real part of $\theta_{1}$, $\theta_{2}$ .

Since the fluctuation is small, Eq.(\ref{eq:24}) is an excellent
approximation. In the same way we also have
\begin{equation}
\begin{split} & \theta_{2}^{(1)}(t)=\theta_{2}^{(0)}(t)+\Delta\theta_{2}^{(1)}(t)\\
 & \Delta\theta_{2}^{(1)}(t)=\int_{0}^{t}dt'\thinspace(\phi_{2}(\theta_{1}^{(0)}(t'),\theta_{2}^{(0)}(t'))-\bar{\phi}_{2})
\end{split}
\label{eq 25}
\end{equation}
where $\bar{\phi}_{2}$ is the average of $\phi_{2}$ over the plane
of the real part of $\theta_{1}$, $\theta_{2}$ . Thus the first
order approximate action is
\begin{equation}
\begin{split} & v_{1}^{(1)}(t)=r_{1}e^{i\theta_{1}^{(1)}(t)}=r_{1}e^{i\theta_{1}^{(0)}(t)+i\Delta\theta_{1}^{(1)}(t)}\\
 & \approx v_{1}^{(0)}(t)(1+i\Delta\theta_{1}^{(1)}(t))\\
 & v_{2}^{(1)}(t)=r_{2}e^{i\theta_{2}^{(1)}(t)}=r_{2}e^{i\theta_{2}^{(0)}(t)+i\Delta\theta_{2}^{(1)}(t)}\\
 & \approx v_{2}^{(0)}(t)(1+i\Delta\theta_{2}^{(1)}(t)),
\end{split}
\label{eq:26}
\end{equation}
We remark that the fluctuation $\Delta\theta_{1}^{(1)}(t),\Delta\theta_{2}^{(1)}(t)$
are complex functions, the real parts represent phase fluctuation,
while the imaginary parts represent the amplitude fluctuation. So
the fluctuation (the deviation from the rigid rotations $v_{1}^{(0)}(t),v_{2}^{(0)}(t)$)
is $\Delta\theta_{1}^{(1)}(t),\Delta\theta_{2}^{(1)}(t)$.

In the domain (a Cantor set \cite{porshel}) where KAM tori persist,
as the power order $n_{s}$ of the square matrix increases, the fluctuation
$\Delta\theta_{1}^{(1)}(t),\Delta\theta_{2}^{(1)}(t)$ approach zero,
and the linear combination coefficients $a_{1},a_{2}$, up to a normalization
constant, approach constants. For a finite $n_{s}$, the standard
deviation of the fluctuation is function of the pair of linear combination
coefficients $\{a_{11},a_{12}\}$,$\{a_{21},a_{22}\}$ in Eq.(\ref{eq:19}).
The most accurate approximation of action-angle variables is obtained
by finding the linear combinations for which the fluctuation is minimized.

For clarity, the zeroth order approximation of trajectory is defined
by $v_{1}^{(0)}\equiv v_{1}(x^{(0)},p_{x}^{(0)},y^{(0)},p_{y}^{(0)})$,
$v_{1}^{(0)}\equiv v_{2}(x^{(0)},p_{x}^{(0)},y^{(0)},p_{y}^{(0)})$,
where $x^{(0)},p_{x}^{(0)},y^{(0)},p_{y}^{(0)}$ are calculated as
function of $\theta_{1}^{(0)},\theta_{2}^{(0)}$, using Eq.(\ref{eq:22})
and the inverse function of Eq.(\ref{eq:19}). In the same way the
first order approximation of trajectory is defined by $v_{1}^{(1)}\equiv v_{1}(x^{(1)},p_{x}^{(1)},y^{(1)},p_{y}^{(1)}),$
$v_{2}^{(1)}\equiv v_{2}(x^{(1)},p_{x}^{(1)},y^{(1)},p_{y}^{(1)})$.
For an exact trajectory, we define $v_{1}^{(e)}\equiv v_{1}(x^{(e)},p_{x}^{(e)},y^{(e)},p_{y}^{(e)}),$
$v_{2}^{(1)}\equiv v_{2}(x^{(e)},p_{x}^{(e)},y^{(e)},p_{y}^{(e)})$.
We refer $x^{(e)},p_{x}^{(e)},y^{(e)},p_{y}^{(e)}$as the exact solution
of Eq.(\ref{eq:I1}), even though we often use the forward integration
to represent them in the following. For each set of linear combinations
$\{a_{11},a_{12}\}$, $\{a_{21},a_{22}\}$, we associate it with the
function $v_{1},v_{2}$ Eq.(\ref{eq:19}), the exact solution $v_{1}^{(e)},v_{2}^{(e)}$,
its zeroth order action (the rigid rotations) $v_{1}^{(0)},v_{2}^{(0)}$,
and the first order solution $v_{1}^{(1)},v_{2}^{(1)}$.

We need to clarify the terminology used here regarding the order of
an approximation. If the zeroth order actions $v_{1}^{(0)},v_{2}^{(0)}$
in Eq.(\ref{eq:22}) are calculated based on square matrix $M$ of
power order $n_{s}$, we call the approximate actions $v_{1}^{(1)},v_{2}^{(1)}$as
the first order perturbation on the zeroth order action based on the
square matrix of power order $n_{s}$. We would like to emphasize
the distinction between the power order of a square matrix, and the
perturbation order in the perturbation theory. We will demonstrate
that if the power order $n_{s}$ is sufficiently high for the Jordan
matrix, the perturbation to the zeroth order $v_{1}^{(0)},v_{2}^{(0)}$
would be small enough that only first order perturbation would result
in much more highly accurate actions than $v_{1}^{(0)},v_{2}^{(0)}$.

Once $v_{1}^{(1)},v_{2}^{(1)}$ in Eq.(\ref{eq:26}) are calculated,
we use the inverse function of Eq.(\ref{eq:19}) to calculate the
more accurate trajectory $x^{(1)},$ $p_{x}^{(1)},$$y^{(1)},p_{y}^{(1)}$,
and apply the method developed in section \ref{sec:CalLin} to find
the linear combinations with minimum fluctuation.

$x,p_{x},y,p_{y}$ are defined as periodic function of $\theta_{1},\theta_{2}$
by the inverse function of Eq.(\ref{eq:19}): its left hand side is
invariant when $\Delta\theta_{1},\Delta\theta_{2}=2\pi$, even with
the analytic continuation of $\theta_{1},\theta_{2}$ in complex domain.
This fact makes it obvious the next step to derive $v_{1}^{(1)},v_{2}^{(1)}$
is to solve the problem by Fourier transform as follows.

\section{\label{sec:IV}Fourier expansion of fluctuation of the action-angle
approximation}

To calculate the fluctuation given by Eq.(\ref{eq:26}), we write
the two dimensional Fourier transform of $\phi_{1}(\theta_{1},\theta_{2}),\phi_{2}(\theta_{1},\theta_{2})$
in Eq.(\ref{eq:21})
\begin{equation}
\begin{split} & \phi_{1}(\theta_{1},\theta_{2})=\sum_{n,m}\widetilde{\phi}_{1nm}e^{in\theta_{1}}e^{im\theta_{2}}\\
 & \phi_{2}(\theta_{1},\theta_{2})=\sum_{n,m}\widetilde{\phi}_{2nm}e^{in\theta_{1}}e^{im\theta_{2}}
\end{split}
\label{eq29}
\end{equation}
In the first order approximation, $\theta_{1},\theta_{2}$ are replaced
by the zeroth order approximation $\theta_{1}^{(0)}(t),\theta_{2}^{(0)}(t)$.
We have the fluctuation (the deviation from a rigid rotation) Eq.(\ref{eq:24})
\begin{equation}
\begin{split} & \Delta\theta_{1}^{(1)}(t)=\int_{0}^{t}dt'\thinspace\sum_{n,m}\widetilde{\phi}_{1nm}e^{in\theta_{1}^{(0)}(t')}e^{im\theta_{2}^{(0)}(t')}-\bar{\phi}_{1}t\\
 & \equiv\thinspace\sum_{n,m}\widetilde{\theta}_{1nm}e^{in\theta_{1}^{(0)}(t)}e^{im\theta_{2}^{(0)}(t)}
\end{split}
\label{eq30}
\end{equation}
Identify the constant $\widetilde{\phi}_{100}$ as the mean phase
shift rate $\bar{\phi}_{1}$, the term linear in t canceled, and $\omega_{1}\equiv\mu+\widetilde{\phi}_{100}$
in $\theta_{1}^{(0)}$. Thus the deviation from rigid rotation is
expressed as a function $\Delta\theta_{1}^{(1)}(\theta_{1},\theta_{2})$
on the plane of angular variables $\theta_{1}^{(0)},\theta_{2}^{(0)}$.
The Fourier expansion coefficients $\widetilde{\theta}_{1nm}$ are
determined as follows. In the integration in Eq.(\ref{eq30}), for
the terms with $|n|+|m|\neq0$, i.e., for all the terms except for
the term with both n and m equal to zero 
\begin{equation}
\begin{split} & \int_{0}^{t}dt'\widetilde{\phi}_{1nm}e^{in\theta_{1}^{(0)}(t')}e^{im\theta_{2}^{(0)}(t')}\\
 & =\widetilde{\phi}_{1nm}e^{i(n\theta_{10}+m\theta_{20})}\int_{0}^{t}dt'e^{i(n\omega_{1}+m\omega_{2})t'}\\
 & =\frac{\widetilde{\phi}_{1nm}}{i(n\omega_{1}+m\omega_{2})}(e^{in\theta_{1}^{(0)}(t)}e^{im\theta_{2}^{(0)}(t)}-e^{i(n\theta_{10}+m\theta_{20})})
\end{split}
\label{eq32}
\end{equation}
where we use the definition of the zeroth order action angle $\theta_{1}^{(0)},\theta_{2}^{(0)}$
Eq.(\ref{eq:22}). Compare Eq.(\ref{eq30}) with Eq.(\ref{eq32}),
we found 
\begin{equation}
\begin{split} & \widetilde{\theta}_{1nm}=\frac{\widetilde{\phi}_{1nm}}{i(n\omega_{1}+m\omega_{2})}\hspace{2cm}for\hspace{2mm}|n|+|m|\neq0\\
 & \widetilde{\theta}_{100}=-\sum_{\tiny{\begin{matrix}n,m\\
|n|+|m|\neq0
\end{matrix}}}\widetilde{\theta}_{1nm}e^{i(n\theta_{10}+m\theta_{20})}
\end{split}
\label{eq33}
\end{equation}
Similar to Eq.(\ref{eq30}) and Eq.(\ref{eq33}), we identify $\widetilde{\phi}_{200}$
as the mean phase shift rate $\bar{\phi}_{2}$, , and $\omega_{2}\equiv\mu+\widetilde{\phi}_{200}$
in $\theta_{2}^{(0)}$.

\begin{equation}
\begin{split} & \Delta\theta_{2}^{(1)}(t)=\thinspace\sum_{n,m}\widetilde{\theta}_{2nm}e^{in\theta_{1}^{(0)}(t)}e^{im\theta_{2}^{(0)}(t)}\\
 & \widetilde{\theta}_{2nm}=\frac{\widetilde{\phi}_{2nm}}{i(n\omega_{1}+m\omega_{2})}\hspace{2cm}for\hspace{2mm}|n|+|m|\neq0\\
 & \widetilde{\theta}_{200}=-\sum_{\tiny{\begin{matrix}n,m\\
|n|+|m|\neq0
\end{matrix}}}\widetilde{\theta}_{2nm}e^{i(n\theta_{10}+m\theta_{20})}
\end{split}
\label{eq34}
\end{equation}
The first order approximation $v_{1}^{(1)}(t),v_{2}^{(1)}(t)$ in
Eq.(\ref{eq:26}) can be written as
\begin{equation}
\begin{split} & v_{1}^{(1)}(t)=r_{1}e^{i\theta_{1}^{(0)}(t)}(1+i\thinspace\sum_{k,m}\widetilde{\theta}_{1km}e^{ik\theta_{1}^{(0)}(t)}e^{im\theta_{2}^{(0)}(t)})\\
 & v_{2}^{(1)}(t)=r_{2}e^{i\theta_{2}^{(0)}(t)}(1+i\thinspace\sum_{k,m}\widetilde{\theta}_{2km}e^{ik\theta_{1}^{(0)}(t)}e^{im\theta_{2}^{(0)}(t)})
\end{split}
\label{eq35}
\end{equation}
Compare with Eq.(\ref{eq:11}), we find for frequency $\omega_{1},\omega_{2}$,
and the fluctuation for frequency $n\omega_{1}+m\omega_{2}$ ($|n|+|m|\neq0)$

\begin{align}
 & \widetilde{v}_{110}=r_{1}(1+i\widetilde{\theta}_{100}),\ \widetilde{v}_{201}=r_{2}(1+i\widetilde{\theta}_{200})\nonumber \\
 & \widetilde{v}_{1km}=ir_{1}\widetilde{\theta}_{1k-1,m}e^{i((k-1)\theta_{10}+m\theta_{20})}\ (|k-1|+[m|\neq0)\label{eq:32}\\
 & \widetilde{v}_{2km}=ir_{2}\widetilde{\theta}_{2k,m-1}e^{i(k\theta_{10}+(m-1)\theta_{20})}\ (|k|+[m-1|\neq0)\nonumber 
\end{align}
for $v_{1}^{(1)},v_{2}^{(1)}$ respectively.

The pair of linear combination coefficients $\{a_{11},a_{12}\}$,
$\{a_{21},a_{22}\}$ in Eq.(\ref{eq:19}) are defined up to two normalization
constants. If we normalize these constants so that the main frequency
components are normalized to one, it would be the same as the normalization
in Eq.(\ref{eq:12}) when we minimize the fluctuation, as given by
$g_{0}$. Notice that when the indices $n$ and $m$ in Eq.(\ref{eq30})
are compared with that in Eq.(\ref{eq:11}) and Eq.(\ref{eq35}),
there is a different meaning of the indices: either $m$, or $n$
would be different by 1..

Since the steady phase advance rates are $\omega_{1}\equiv\mu+\bar{\phi}_{1}$,
$\omega_{2}\equiv\mu+\bar{\phi}_{2}$, stable motion requires $\text{{Im}}\widetilde{\phi}_{100}=0$
and $\text{{Im}}\widetilde{\phi}_{200}=0$. In the iteration process
to find solution, as will be explained in the following sections,
they converge to zero very fast, and serve as a test for the convergence.
Numerical study shows that taking $\omega_{1}\equiv\mu+\text{{Re}(}\bar{\phi}_{1})$,
$\omega_{2}\equiv\mu+\text{{Re}(}\bar{\phi}_{2})$ in the iteration
can make the convergence slightly faster in the initial few steps
of the iteration.

\section{\label{sec:Iteration}Iteration to improve precision}

\subsection{\label{sub:Iteration:-principle-and}Iteration: principle and procedure}

If the zeroth order approximation, the rigid rotations $v_{1}^{(0)},v_{2}^{(0)}$
in Eq.(\ref{eq:22}) determined by $\{a_{11},a_{12}\}$,$\{a_{21},a_{22}\}$
in Eq.(\ref{eq:19}), are sufficiently close to the persistent invariant
KAM tori, the perturbation in Section \ref{sec:IIIPerturbation} and
\ref{sec:IV} would be small: $|\widetilde{\theta}_{lnm}|\ll1$ for
$|n|+|m|\neq0$ and $l=\{1,2\}$, the first order approximation $v_{1}^{(1)},v_{2}^{(1)}$
in Eq.(\ref{eq35}) would provide more accurate solution. When the
fluctuation of the more accurate solution is minimized using the method
developed in Section \ref{sec:CalLin}, the further optimized $\{a_{11},a_{12}\}$,$\{a_{21},a_{22}\}$
in Eq.(\ref{eq:19}) would correspond to a new set of rigid rotations
$v_{1}^{(0)},v_{2}^{(0)}$ more close to the KAM tori. There are two
steps here: 1. find $v_{1}^{(1)},v_{2}^{(1)}$ so they are more close
to the exact solution $v_{1}^{(e)},v_{2}^{(e)}$; 2. find new $\{a_{11},a_{12}\}$,$\{a_{21},a_{22}\}$
so $v_{1}^{(1)},v_{2}^{(1)}$ are more close to the rigid rotations
$v_{1}^{(0)},v_{2}^{(0)}$.

Hence the solution to the perturbation theory developed in Section
3-5 requires an iteration process. Starting from a first trial linear
combination coefficients, we calculate zeroth order approximate action-angle
variables $v_{1}^{(0)},v_{2}^{(0)}$. These are used to calculate
$v_{1}^{(1)},v_{2}^{(1)}$, which is more close to $v_{1}^{(e)},v_{2}^{(e)}$.
Then $v_{1}^{(1)},v_{2}^{(1)}$ are used to find the trajectory $x^{(1)}$,$p_{x}^{(1)}$,
$y^{(1)}$,$p_{y}^{(1)}$ by the inverse function of Eq.(\ref{eq:19}).
This in turn leads to the left eigenvectors $w_{j}$ in Eq.(\ref{eq:10}).
Then, a more accurate linear combination coefficients are derived
by minimizing the fluctuation in Section \ref{sec:CalLin}. Since
the new linear combinations give a new set of $v_{1}^{(1)},v_{2}^{(1)}$
for the same trajectory $x^{(1)}$,$p_{x}^{(1)}$, $y^{(1)}$,$p_{y}^{(1)}$
but with minimized fluctuation, the rigid rotations $v_{1}^{(0)},v_{2}^{(0)}$
represented by the new linear combination in the second iteration
also represent $v_{1}^{(1)},v_{2}^{(1)}$ of previous iteration very
well. Hence we can repeat the iteration to calculate $v_{1}^{(1)},v_{2}^{(1)}$
for the second iteration.

For this procedure, the main issues are: 
\begin{enumerate}
\item How to find the initial linear combinations? Or the first zeroth order
approximation $v_{1}^{(0)},v_{2}^{(0)}$? 
\item What is the region of convergence for the iteration procedure? What
is its relation to the chaotic boundary? 
\end{enumerate}
As for issue 1, as explained in Section \ref{sec:CalLin}, if we have
a forward numerical integration of the dynamic equations, we can use
Fourier expansion to determine the initial trial linear combinations.
However, since our goal is to solve the problem without forward numerical
integration, we shall use the Henon-Heiles problem as an example to
show that in much of the region where the invariant KAM tori persist,
the solution based on differential equations at the initial position
determined from Eq.(\ref{14}) can be used as the initial trial linear
combinations, and the iteration leads to convergent result. This would
not always lead to convergence. However, the linear combinations for
a solution of a smaller amplitude case can be used as initial trial
linear combinations for the solution of a larger amplitude case. Numerical
examples show that this approach is valid in general, when we know
the solution near the fixed points.

About issue 2, if the initial trial action-angle variables $v_{1}^{(1)},v_{2}^{(1)}$
are dominated by $\omega_{1}$, $\omega_{2}$ components respectively,
i.e., they are very close to rigid rotations, the perturbation terms
in Eq.(\ref{eq:21}) would be very small, and iteration would converge
very fast.

When we use the differential equations at the initial position to
determine the initial trial linear combinations from Eq.(\ref{14}),
for large amplitude case, when $v_{2}^{(1)}$ is not dominated by
$\omega_{2}$ components, the first few iterations progress slowly.
After a few iterations, the main frequency components $\omega_{1}$,
$\omega_{2}$ establish dominance in $v_{1},v_{2}$ respectively,
the iteration converges very fast. We then increase the number $n_{v}$
of left eigenvectors to improve the precision. When we found convergence,
the result is always in highly accurate agreement with the forward
integration unless it is very close to the stability boundary.

Numerical examples show that for small amplitude the iteration converges
fast. We can use the result of a previous convergence as the trial
solution when we further increase the amplitude. When the amplitude
increases, the number of iterations to reach convergence increases.
When close to the separatrix the residual fluctuation increases, we
need to reduce the step size, and we may need to further increase
$n_{v}$. Eventually the iteration is no longer convergent when very
close to the stability boundary.

But our knowledge about the convergence so far is limited to numerical
study. It is an open question whether when the KAM tori persist it
is always possible to find a way to keep the iteration converges by
increasing $n_{s},n_{v}$ and by reducing the step size to reach an
increased amplitude. The relation between the convergence and the
chaotic boundary is still unknown, and remains to be very important
open issue, even though whether the iteration converges already provides
information about the stability boundary.

In the following, we shall give examples showing the minimization
leads to highly accurate action-angle variables.

\begin{figure}[t]
\includegraphics[width=0.48\textwidth]{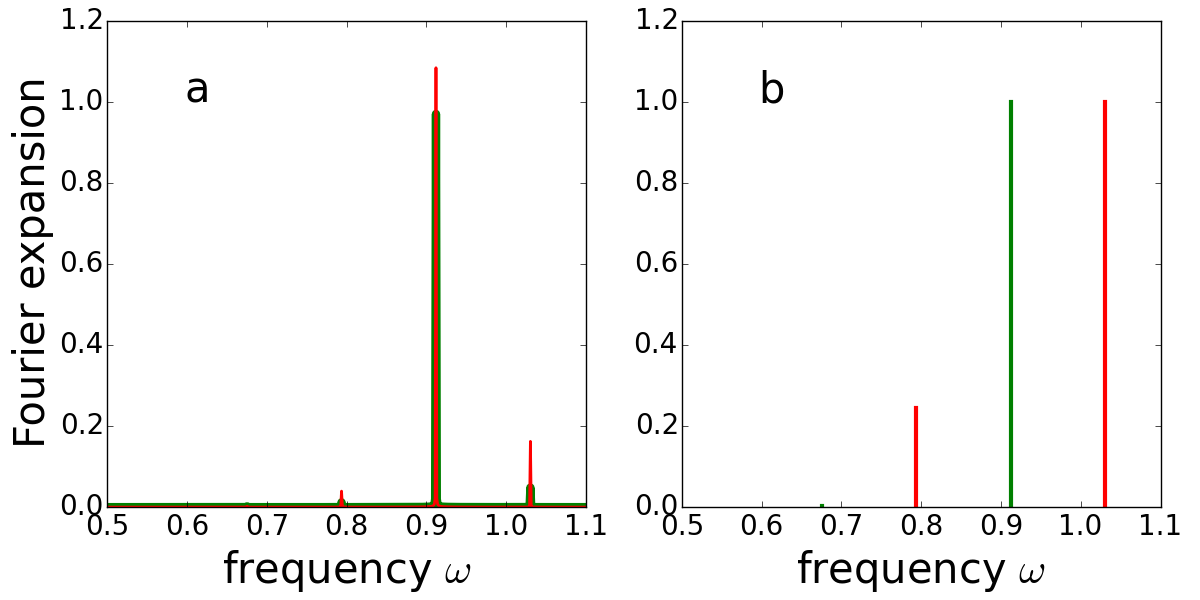}\vspace{-1.0em}

\caption{Fourier expansion vs. frequency for $x_{0},y_{0},p_{y0}=0,0,0.18$,
power order $n_{s}=5$, a:$w_{x0}$ (green) $w_{y0}$ (red) b:$v_{1}$
(green),$v_{2}$ (red). $w_{x0},w_{y0}$ are normalized so their initial
values are 1. $w_{x0}$ (green) and $w_{y0}$ (red) are nearly parallel
except $w_{y0}$ has a second lower peak at $\omega_{2}=1.03$ with
$|\widetilde{w}_{y001}|=0.164$, higher than the second peak for $w_{x0}$,
$|\widetilde{w}_{x001}|=0.049$. Hence the two vectors are independent,
so it is possible to choose a set of vector basis ($v_{1}$, $v_{2}$)
that they are nearly orthogonal in the frequency space. This is carried
out as shown in Section \ref{sec:CalLin} to calculate $\{a_{11},a_{12}\}$,$\{a_{21},a_{22}\}$
in Eq.(\ref{eq:15}), with the sum over $n$, $m$ in Eq.(\ref{eq:10})
limited to only over the 6 lines in Table I. The result is shown in
b, with $|\widetilde{v}_{1nm}|$ (green) and $|\widetilde{v}_{2nm}|$
(red).}
\vspace{-1.0em}
\end{figure}

\begin{table}[b]
\begin{tabular}{|c|c|c|c|c|}
\hline 
n  & m  & $\omega=n\omega_{1}+m\omega_{2}$  & $|\widetilde{w}_{x0nm}|$  & $|\widetilde{w}_{y0nm}|$\tabularnewline
\hline 
\hline 
1  & 0  & 0.9120  & 0.972  & 1.089\tabularnewline
\hline 
0  & 1  & 1.0303  & 0.049  & 0.164\tabularnewline
\hline 
2  & -1  & 0.7937  & 0.013  & 0.043\tabularnewline
\hline 
0  & 0  & 0.0000  & 0.012  & 0.014\tabularnewline
\hline 
2 & 0 & 1.8240  & 0.006  & 0.006\tabularnewline
\hline 
3  & -1  & 1.7057 & 0.000  & 0.005\tabularnewline
\hline 
\end{tabular}

\caption{Absolute value of Fourier expansion coefficients}
\end{table}

\subsection{\label{sub:Numerical-Examples}Numerical Examples}

Since our goal is to find the solution without using the numerical
forward integration, we shall illustrate the iteration procedure to
find the action-angle variables using an example Eq.(\ref{eq:I1})
with initial condition $E=H=1/12\approx.083$3, $x_{0},y_{0},p_{y0}=$
$0,0,0.18$ ($p_{x0}$ is determined from $H$ in Eq.(\ref{eq:I1})).
However, in order to check the result, we need to compare the iteration
progress and the result with the forward integration. Hence we first
present the result obtained by forward integration, represented by
a trajectory $x^{(e)},p_{x}^{(e)},y^{(e)},p_{y}^{(e)}$.

Following Section \ref{sec:CalLin}, we search for two linear combinations
for which $v_{1}$, $v_{2}$ nearly represent two independent rigid
rotations, i.e., with two different frequencies and with minimized
fluctuation. We take $n_{s}=5,n_{v}=2$. The Fourier transform of
$w_{x0},w_{y0}$, with the trajectory $x^{(e)},p_{x}^{(e)},y^{(e)},p_{y}^{(e)}$
is shown in Fig.1a. In Table 1 we list the top 6 peaks with their
frequencies, indices ${n,m}$, and peak heights. The spectrum of $v_{1}$,
$v_{2}$ with fluctuation minimized is shown in Fig.1b. The fluctuation
of $|v_{2}|$ mainly due to the line at $\omega=0.7937$ is to be
further reduced by the procedure prescribed in Section \ref{sub:Iteration:-principle-and}. 

But our goal is to find the solution without the forward integration,
so we shall not use the linear combination obtained this way. Instead,
in the following example, we start the iteration procedure from the
initial trial linear combinations $\{a_{11},a_{12}\}$,$\{a_{21},a_{22}\}$
given by Eq.(\ref{14}) derived from the differential equations at
the initial position. And, we introduce normalized action-angle variables
$\bar{v}_{1}\equiv v_{1}/r_{1},\bar{v}_{2}\equiv v_{2}/r_{2}$, so
in many of the following plots of spectrum the fluctuation is relative
to $|\bar{v}_{1}|\approx1,|\bar{v}_{2}|\approx1$ (see Eq.(\ref{eq35})).

\begin{figure}[t]
\includegraphics[width=0.29\textwidth]{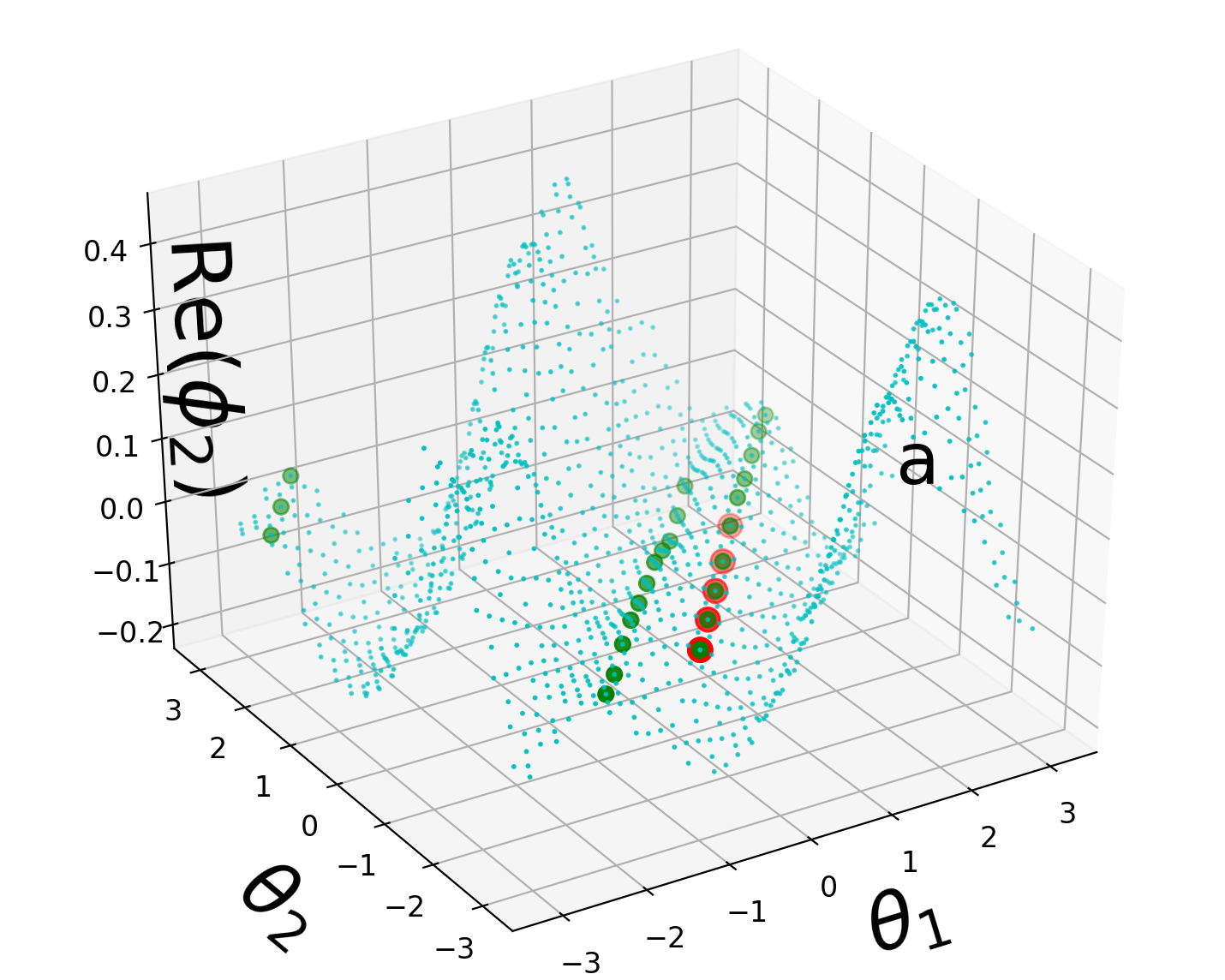}\includegraphics[width=0.19\textwidth]{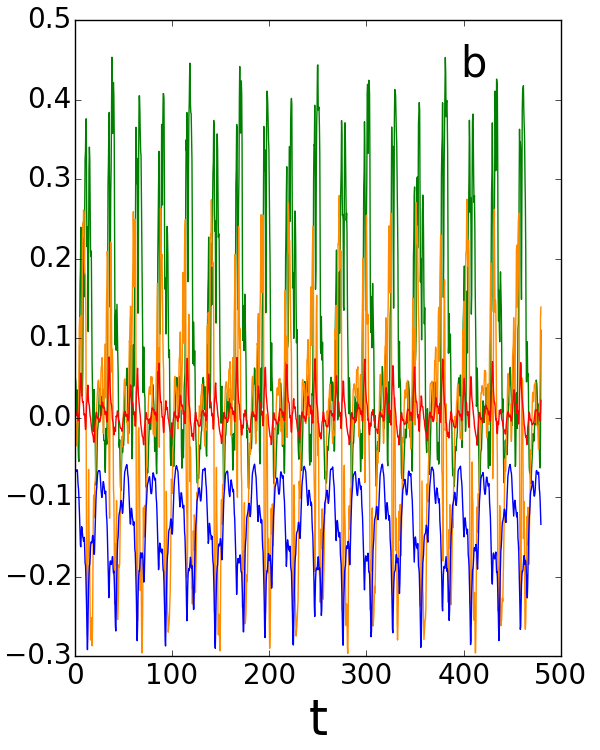}\vspace{-1.0em}

\caption{For $x_{0},y_{0},p_{y0}=0,0,0.18$, linear combinations based on differential
equations at the initial position, a: $\text{Re(}\phi_{2})$ calculated
from Eq.(\ref{eq:21}) as the rigid rotations $v_{1}^{(0)}$, $v_{2}^{(0)}$
in Eq.(\ref{eq:22}) pass through the $\theta_{1}^{(0)}$, $\theta_{2}^{(0)}$
plane. The first 5 points (red), and first 24 points (green) show
how the trajectory scans through the plane. b: $\text{{Re}}\phi_{1}$
(blue), $\text{{Im}}\phi_{1}$(red), $\text{{Re}}\phi_{2}$(green),
$\text{{Im}}\phi_{2}$ (orange) as function of $t$.}
\end{figure}

\begin{figure}[t]
\vspace{-0.5em}\includegraphics[width=0.267\textwidth]{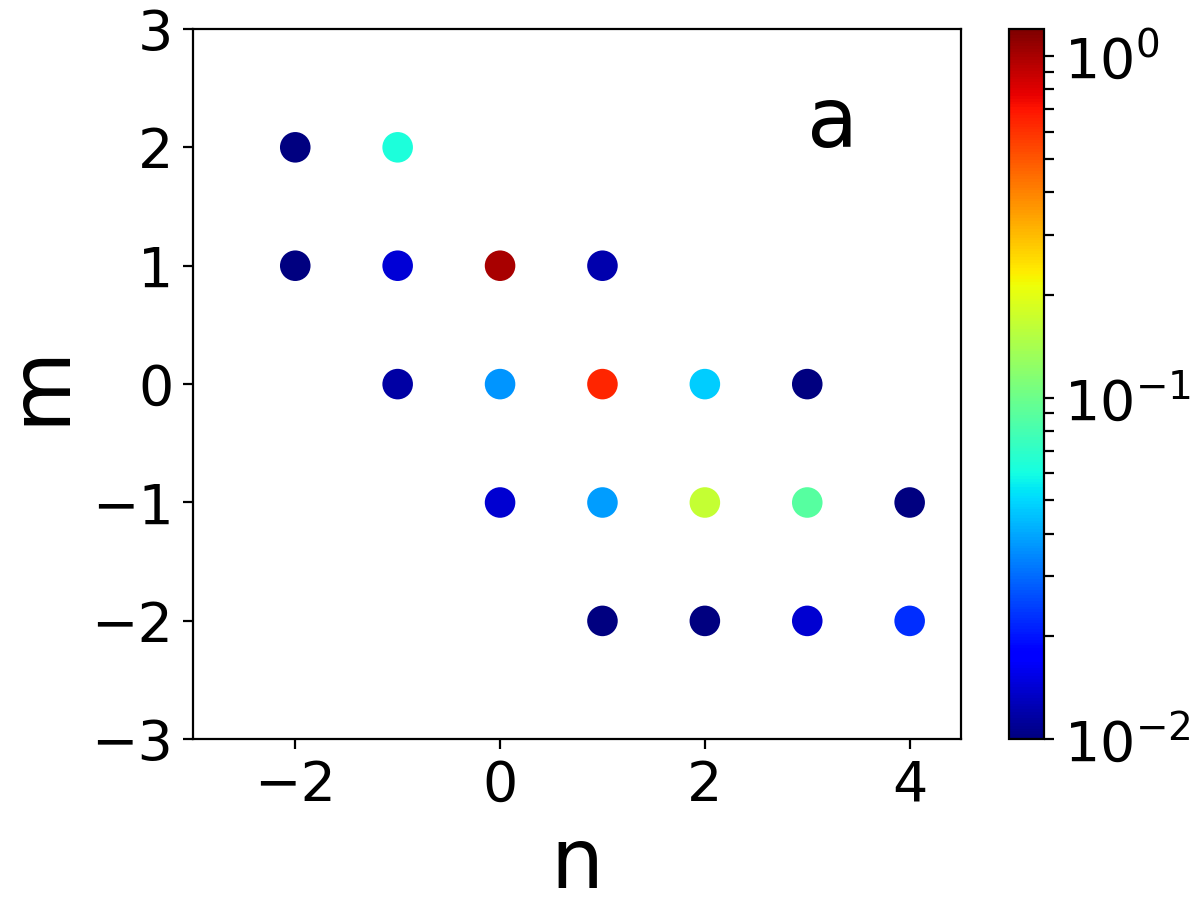}\includegraphics[width=0.213\textwidth]{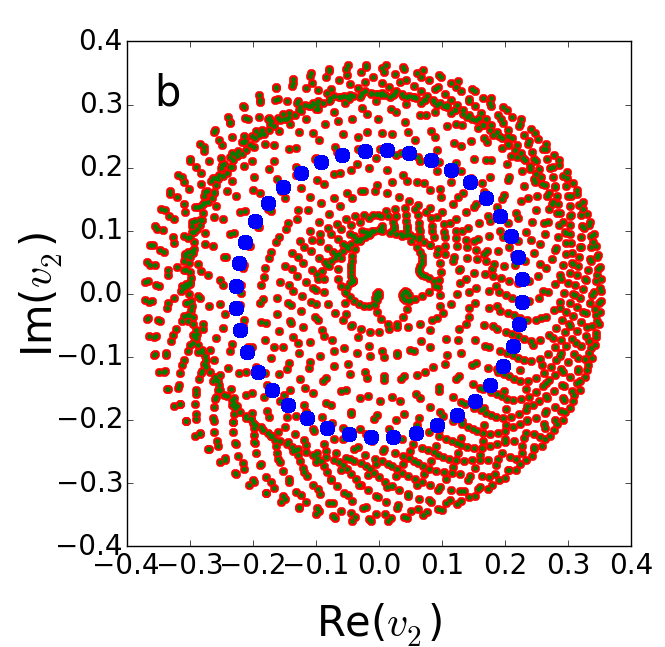}\vspace{-1.0em}

\caption{a. Fourier expansion vs. indices of $\bar{v}_{2}^{(1)}$ in iteration
1, calculated by Eq.(\ref{eq34}). The red dot at $n,m=1,0$ in Fig.3a
corresponds to $\widetilde{\theta}_{21,-1}$ in Eq.(\ref{eq35}).
Notice the top 3 peaks $n,m=$\{0,1\},\{1,0\}\{2,-1\} in Fig.3a correspond
to the top 3 peaks with intensity of 1,0.64, 0.17 in Fig.4b at $\omega_{2}=1.0984$,
$\omega_{1}=0.8600,$ $2\omega_{1}-\omega_{2}=0.6215$ respectively.
b: phase space $v_{2}^{(0)}$(blue), $v_{2}^{(1)}$ (red, calculated
from Eq.(\ref{eq35}) ), $v_{2}^{(1)}$(green): from $v_{1}^{(1)}$,
$v_{2}^{(1)}$, the inverse function of Eq.(\ref{eq:19}) are used
to find the trajectory $x^{(1)},p_{x}^{(1)},y^{(1)},p_{y}^{(1)}$,
which in turn are used to calculate $w_{x0},w_{y0}$ and their Fourier
transform $\widetilde{w}_{jnm}$, and $\widetilde{v}_{lnm}$ in Eq.(\ref{eq:11})
of Section \ref{sec:CalLin}, using the initial trial linear combinations
$\{a_{11},a_{12}\}$,$\{a_{21},a_{22}\}$ of iteration 1. The results
are used to calculate $v_{1}^{(1)}$, $v_{2}^{(1)}$ again and plotted
in b as green dots for a double check because they should coincide
with the red dots. }
\end{figure}

\begin{figure}[t]
\vspace{-1.0em}\includegraphics[width=0.48\textwidth]{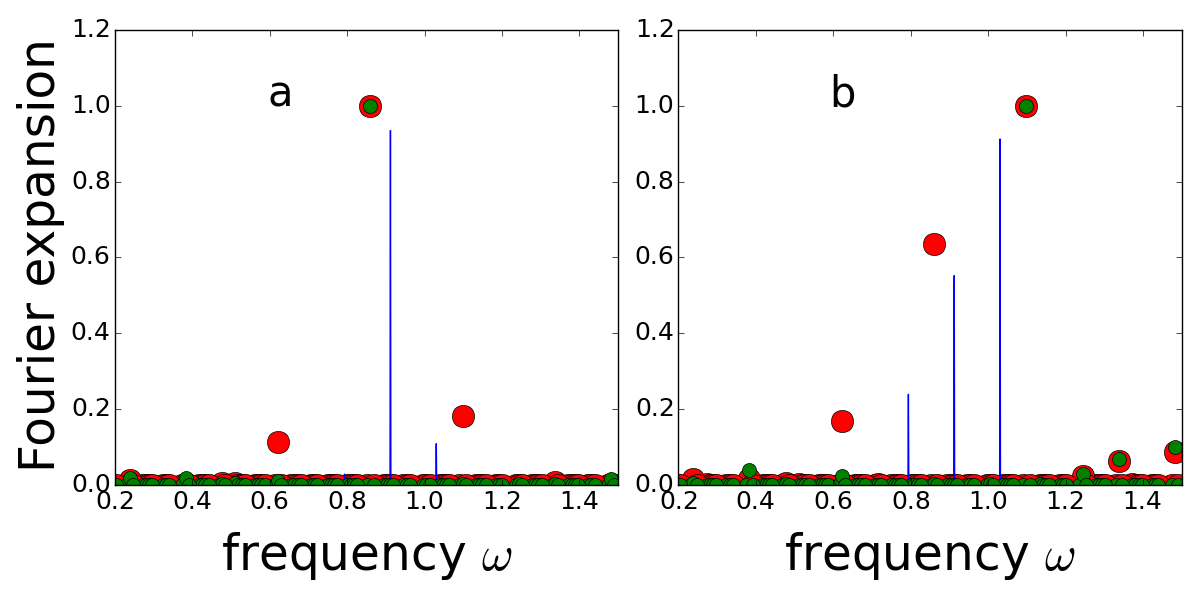}\vspace{-1.0em}

\caption{Fourier expansion vs. frequency, iteration 1, a:$\bar{v}_{1}$ b:
$\bar{v}_{2}$. Red dots are $\widetilde{\theta}_{1nm}$ $\widetilde{\theta}_{2nm}$
of Eq.(\ref{eq33}), Eq.(\ref{eq34}), green dots are $\widetilde{v}_{lnm}$
in Eq.(\ref{eq:11}) using linear combinations Eq.(\ref{eq:15}) for
minimized fluctuation in iteration 1, i.e., the initial linear combination
of iteration 2; blue lines are $\bar{v}_{2}^{(e)}$ in iteration 1.
For the red dots, the intensity 1 at $\omega_{2}=1.098$ , the dominating
line of $\bar{v}_{2}^{(1)}$, is not much more than the intensity
0.64 at $\omega_{1}=0.8600$ while our goal is to find $\bar{v}_{2}^{(1)}$
dominated by the line at $\omega_{2}$. So $\bar{v}_{2}^{(1)}$ can
hardly represent a perturbation to the rigid rotation. The fact $\omega_{1},\omega_{2}=0.860,1.098$
is different from $\omega_{1},\omega_{2}=0.912,1.030$ obtained from
the forward integration as given in Table I is because the errors
in the initial trial linear combination, and the perturbation calculation.
As the iteration progresses, $\omega_{1},\omega_{2}$ approaches $0.912,1.030$.
The frequencies of the red dots $\bar{v}_{2}^{(1)}$ have large errors
relative to the blue lines $\bar{v}_{2}^{(e)}$ showing $\bar{v}_{2}^{(1)}$
is a poor approximation. The green dots are dominated by $\omega_{1}$
and $\omega_{2}$ in Fig.4a and Fig.4b respectively. The rigid rotations
determined by the new linear combinations should be single lines at
$\omega_{1}$ and $\omega_{2}$ in Fig.4a and 4b, hence the rigid
rotation is a good approximation of green dots $\bar{v}_{2}^{(1)}$,
expressed in terms of the new linear combination derived in iteration
1. }
\end{figure}
\begin{figure}[t]
\vspace{-1.0em}\includegraphics[width=0.48\textwidth]{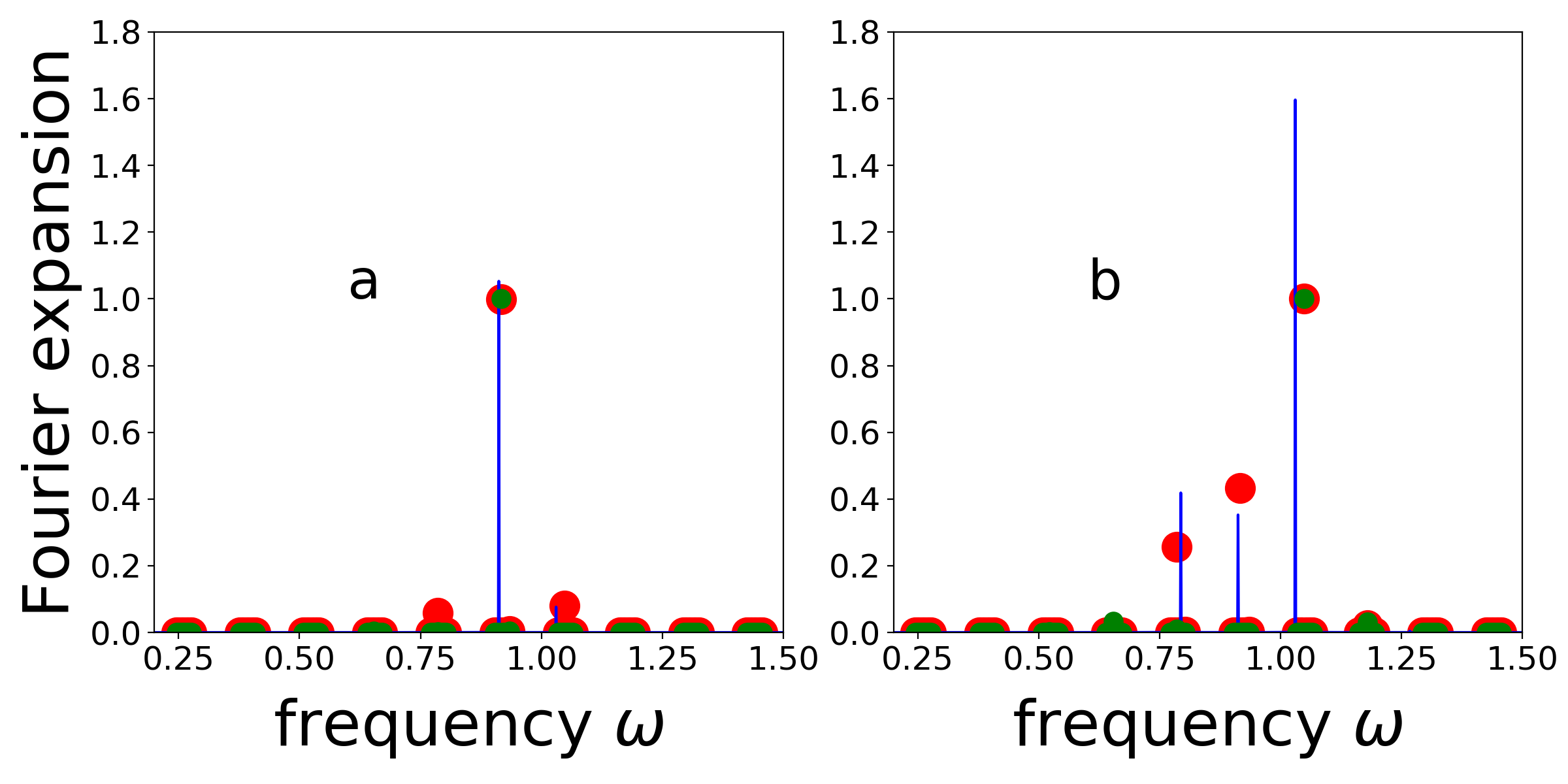}\vspace{-1.0em}\caption{Fourier expansion vs. frequency, iteration 2, a:$\bar{v}_{1}$ b:$\bar{v}_{2}$.
$v_{1}^{(1)}$, $v_{2}^{(1)}$ (red) are calculated from $v_{1}^{(0)}$,
$v_{2}^{(0)}$ by steps prescribed in Section \ref{sec:IIIPerturbation}
and \ref{sec:IV} based on the linear combinations $\{a_{11},a_{12}\}$,$\{a_{21},a_{22}\}$
derived in iteration 1. Again the blue lines are $\bar{v}_{1}^{(e)}$
, $\bar{v}_{2}^{(e)}$. Now the blue line at $\omega_{2}=1.03$ is
much higher than $\omega_{1}=0.912$ in Fig.5b showing that in iteration
2, the rigid rotation $v_{2}^{(0)}$ (a single line at $\omega_{2}$)
is a better approximation than in iteration 1 when compared with Fig.4b,
even though it is still a poor approximation because the line at $\omega_{1}$
is still more than 25\% of the line at $\omega_{2}$. The red dot
at $\omega_{1}=0.917$ in Fig.5b is still 40\% of $\omega_{2}=1.048$,
showing $\bar{v}_{2}^{(1)}$ is still a poor approximation of the
accurate solution $\bar{v}_{2}^{(e)}$ (the blue lines). Clearly this
is due to the fact $v_{2}^{(0)}$ in iteration 2 is still a poor approximation.
But the frequencies of the red dots are much more closer to the blue
lines than Fig.4. Again the green dots show a new set of linear combinations
with minimized fluctuation for $\bar{v}_{1}^{(1)}$, $\bar{v}_{2}^{(1)}$
in iteration 2 , which now is the zeroth order approximation $\bar{v}_{1}^{(0)}$,
$\bar{v}_{2}^{(0)}$ in iteration 3. }
\end{figure}

Then we apply the perturbation theory of Section \ref{sec:IIIPerturbation}
to calculate the zeroth order $v_{1}^{(0)}$, $v_{2}^{(0)}$, and
$\phi_{1}$, $\phi_{2}$ in Eq.(\ref{eq:21}). A comparison of Fig.2a
with Fig.2b helps to understand how the trajectory passes the $\theta_{1}^{(0)}$,
$\theta_{2}^{(0)}$ plane.

\begin{figure}[t]
\vspace{-0.0em}\includegraphics[width=0.48\textwidth]{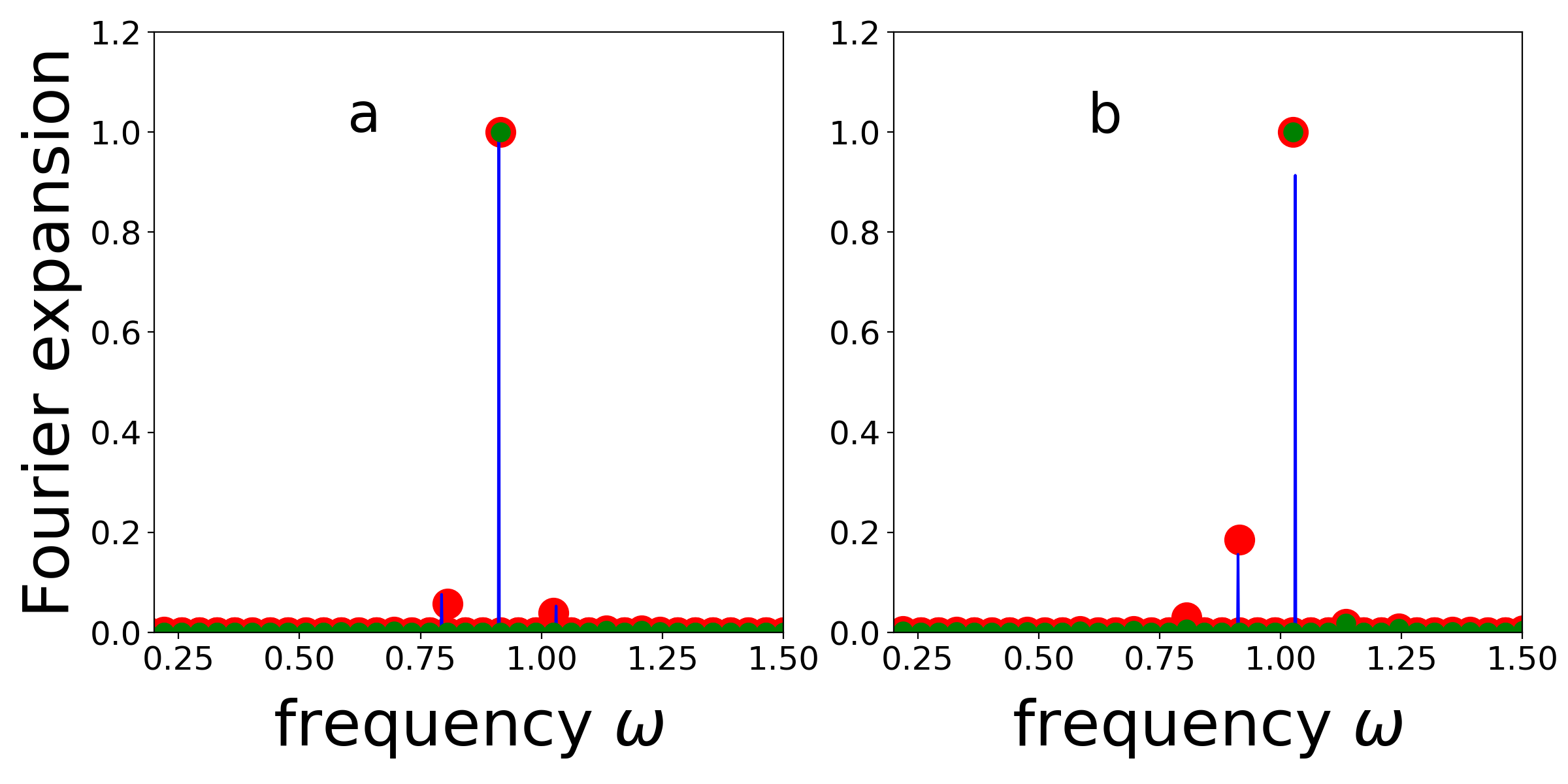}\vspace{-1.0em}\caption{Fourier expansion vs. frequency, iteration 3, a:$\bar{v}_{1}$ b:$\bar{v}_{2}$.
The Fourier expansion of $\bar{v}_{1}^{(1)}$, $\bar{v}_{2}^{(1)}$
(red dots) of iteration 3 show they are dominated by red dots $\omega_{1}$
and $\omega_{2}$ respectively. The fluctuation in $|\bar{v}_{2}^{(1)}|$
is the line at $\omega_{1}=0.915$ with height 0.19, much smaller
the main line at $\omega_{2}=1.025$ with height 1. All the red, green
dots and blue lines are dominated by $\omega_{1}$ and $\omega_{2}$
respectively in Fig.6a,b, showing the iteration is converging.}
\end{figure}
\begin{figure}[t]
\vspace{-1.0em}\includegraphics[width=0.25\textwidth]{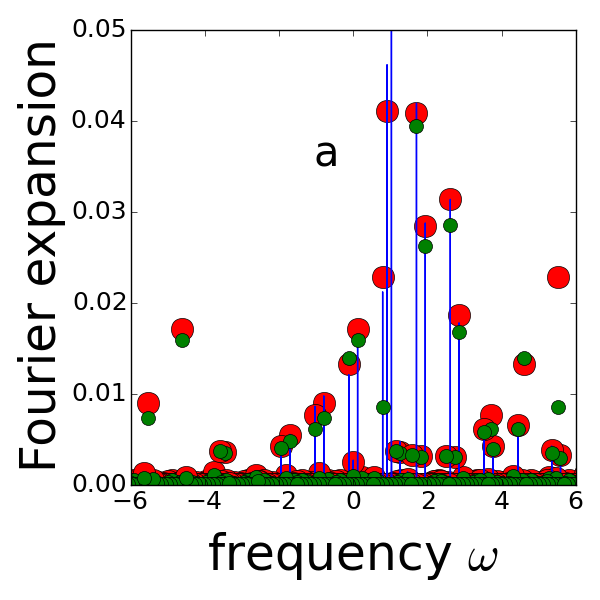}\includegraphics[width=0.23\textwidth,height=0.25\textwidth]{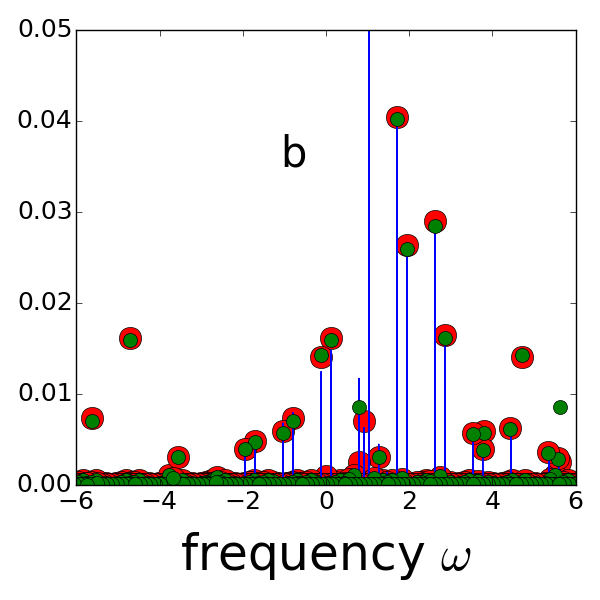}\vspace{-1.0em}

\caption{Fourier expansion vs. frequency. a:$\bar{v}_{2}$ iteration 4, b:
$\bar{v}_{2}$ iteration 5}
\end{figure}

\begin{figure}[t]
\vspace{-0.0em}\includegraphics[width=0.25\textwidth]{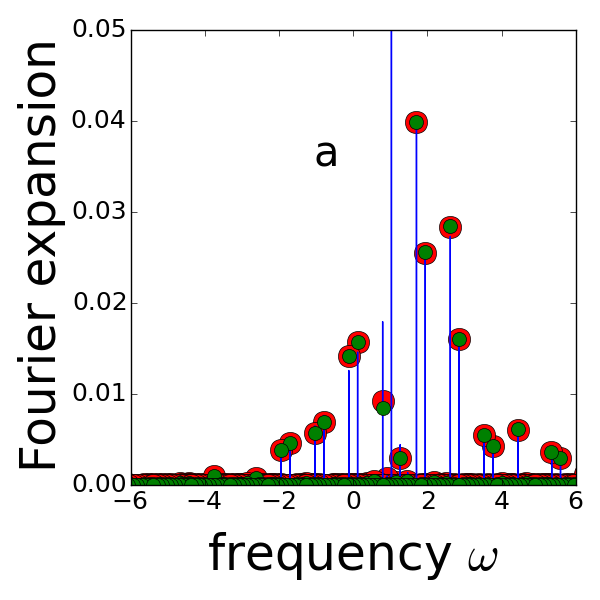}\includegraphics[width=0.23\textwidth,height=0.25\textwidth]{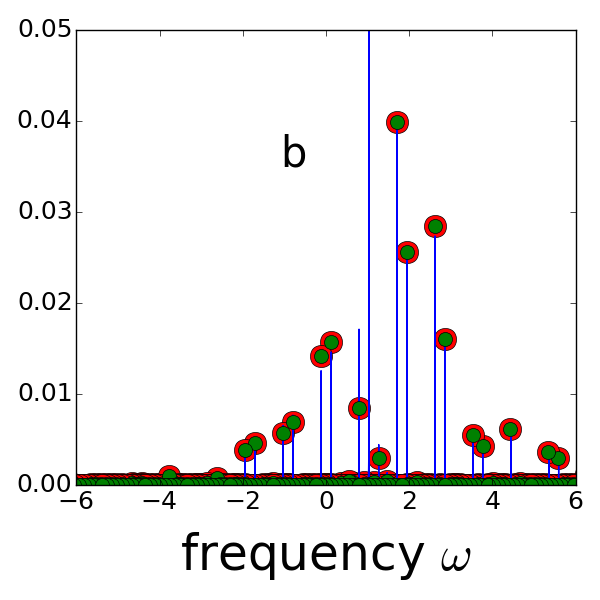}\vspace{-1.0em}

\caption{Fourier expansion vs. frequency. a :$\bar{v}_{2}$ iteration 6, b:
$\bar{v}_{2}$ iteration 7. The fluctuation (about 4\%) is calculated
to be in agreement with the forward integration, i.e.,$\bar{v}_{1}^{(1)}\approx\bar{v}_{1}^{(e)},\bar{v}_{2}^{(1)}\approx\bar{v}_{2}^{(e)}$,
accurate to about 0.6\%. This is visible when compare the red dots
(the first order approximation based on the rigid rotations) with
the blue lines (the forward integration). The red dots almost coincide
with the green dots (the new linear combinations for $v_{2}^{(1)}$at
the end of iteration 7) indicates convergence of the iteration.}
\end{figure}

If we increase the left eigenvectors to $w_{x0},w_{y0},w_{x1},w_{y1}$
$(n_{v}=4)$, we can obtained $|v_{2}|$ with less fluctuation. However,
because the vectors $w_{x1},w_{y1}$ are also dominated by $\omega_{1}$,
$\omega_{2}$, components of frequency other than $\omega_{1}$, $\omega_{2}$
are small. The inclusion of $w_{x1},w_{y1}$ in the minimization process
often results in large contribution from $w_{x1},w_{y1}$ to suppress
other components, even if the fluctuation is small, thus makes the
iteration convergence slow or even stopped before the dominance of
the main frequency components $\omega_{1}$, $\omega_{2}$ in $v_{1},v_{2}$
is established respectively. Hence we take $n_{v}=2$ at first so
the iteration converges quickly.

The 2-D Fourier transform of $\bar{v}_{2}^{(1)}$ vs. indices based
on the steps in Section \ref{sec:IV} in iteration 1 is shown in Fig.3a,
and, as function of frequency, is shown in Fig.4b as the red dots.
In Fig.3b we show $v_{2}^{(1)}$ in the phase space of $\text{Re}(v_{2})$,
$\text{{Im}}(v_{2})$ (red dots). The blue dots are $v_{2}^{(0)}$.
Fig.3b shows the deviation of $v_{2}^{(1)}$ from $v_{2}^{(0)}$ is
so large that it is hardly a perturbation to the rigid rotation. This
is observed in Fig.4b too. In Fig.4b, the frequencies of the red dots
$\bar{v}_{2}^{(1)}$ have large errors relative to the blue lines
$\bar{v}_{2}^{(e)}$also showing it is a very poor approximation.

With this provision, the next step is to find the new linear combinations
$\{a_{11},a_{12}\}$,$\{a_{21},a_{22}\}$ by the minimization procedure
from Eq.(\ref{eq:12}) to Eq.(\ref{eq:15}). The spectrum of $\bar{v}_{1}^{(1)}$,
$\bar{v}_{2}^{(1)}$ obtained by the new linear combination of iteration
1 is shown as green dots in Fig.4a and Fig.4b, they are close to rigid
rotations. Therefore the rigid rotations represented by the new linear
combinations serves as the zeroth order approximation in iteration
2. This completes the iteration 1.

The iteration 2 follows the same steps. Now the zeroth order approximation
$v_{1}^{(0)}$, $v_{2}^{(0)}$, calculated from the linear combinations
derived in iteration 1, and represent single lines at $\omega_{1}$
and $\omega_{2}$ respectively in Fig.5a, 5b, are close to the first
order approximation in iteration 1. Fig.5b shows that in iteration
2, the rigid rotation $v_{2}^{(0)}$ is a better approximation than
in iteration 1 even though it is still a poor approximation. Clearly
because of this, Fig.5b also shows $\bar{v}_{2}^{(1)}$(red) is still
a poor approximation of the accurate solution (blue). 

The spectrum in Fig.6a,b shows $v_{1}^{(1)}$, $v_{2}^{(1)}$ in iteration
3 are well represented by rigid rotations. To reduce the fluctuation
due to other harmonics such as $2\omega_{1}-\omega_{2}=0.805$ in
iteration 3, we repeat the minimization procedure of Section \ref{sec:CalLin}
in iteration 2 with increased number of left eigenvectors $n_{v}=4$
in Eq.(\ref{eq:11}). The green dots in Fig.5a,b and the plots of
iteration 3 in Fig.6a,b are all based on linear combinations calculated
with $n_{v}=4$. Fig.6a,b shows the iteration is converging.

Since the $v_{1}^{(1)}$ always has much smaller fluctuation than
$v_{2}^{(1)}$, in the following iterations we only show the spectrum
of $v_{2}^{(1)}$. Because the fluctuation decreases rapidly to lower
than a few percent of the main peaks of height 1, we change the axis
scale to 5\% maximum, and increased the frequency range to show wider
noise spectral range. These plots in iteration 4-7 are shown in Fig.7-8,
showing the fluctuation spectrum pattern converges. Consider the very
complicated noise spectrum in Fig.8b, the extremely detailed agreement
between the red dots (square matrix solution) and the blue lines (the
forward integration) is very pronounced. 

In Fig.8b the final convergent result $v_{1}^{(e)}$, $v_{2}^{(e)}$
have fluctuation over the rigid rotations of about 4\%. We see $v_{1}^{(1)}\approx v_{1}^{(e)}$,
$v_{2}^{(1)}\approx$ $v_{2}^{(e)}$, 
\begin{figure}[b]
\includegraphics[width=0.5\columnwidth]{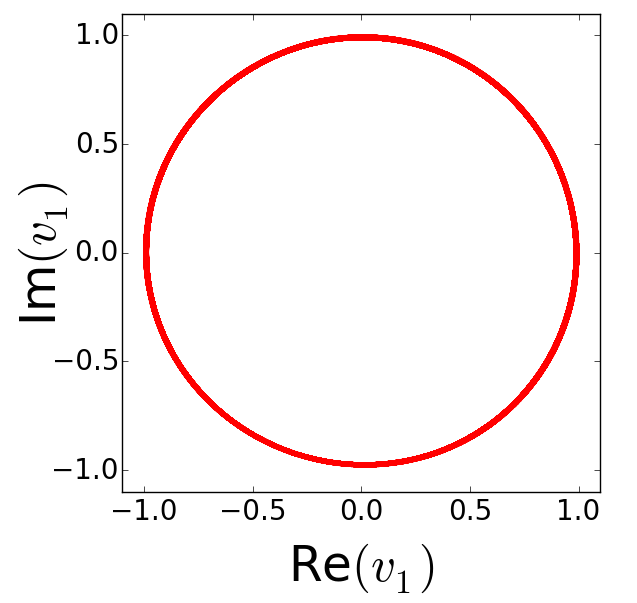}\includegraphics[width=0.5\columnwidth]{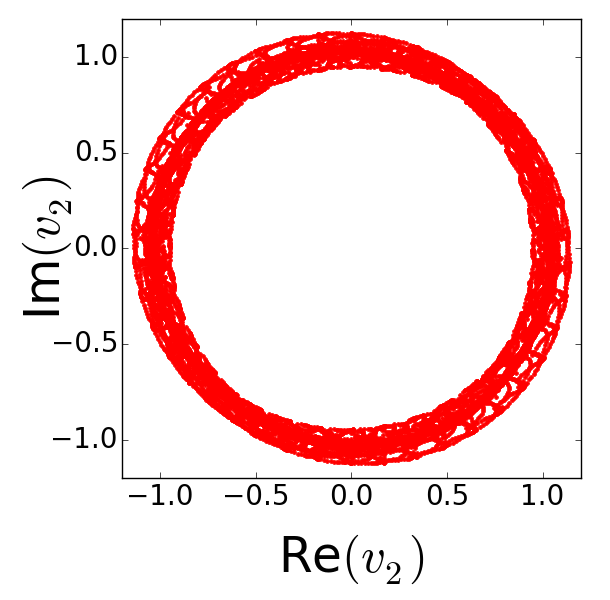}\vspace{-1.0em}

\caption{left:$v_{1}^{(e)}$ , right:$v_{2}^{(e)}$ iteration 7. $v_{1}^{(e)}$
is nearly a perfect circle, $v_{2}^{(e)}$ has a fluctuation of the
radius about 4\%, as expected from the spectrum in Fig.8b. }
\end{figure}
accurate to about 0.6\%, much less than 4\%. This observation allows
us to derive a set of much more accurate KAM invariants $\bar{v}_{01}$,$\bar{v}_{02}$,
as will be explained in the Section \ref{sec:Transform-Rotation-with}.

Result of iteration: Fig.9 show $v_{1}^{(e)}$, $v_{2}^{(e)}$ of
the last linear combination. Fig.10 shows $v_{2}^{(1)}$, in agreement
with $v_{2}^{(e)}$ in Fig.9. Fig.11 shows the upper half $y,p_{y}$
plane of the Poincare surface section with x crosses zero, showing
the result from square matrix gives much better agreement with forward
integration.\begin{wrapfigure}{o}{0.5\columnwidth}%
\begin{raggedright}
\vspace{-1.0em}\includegraphics[width=4cm]{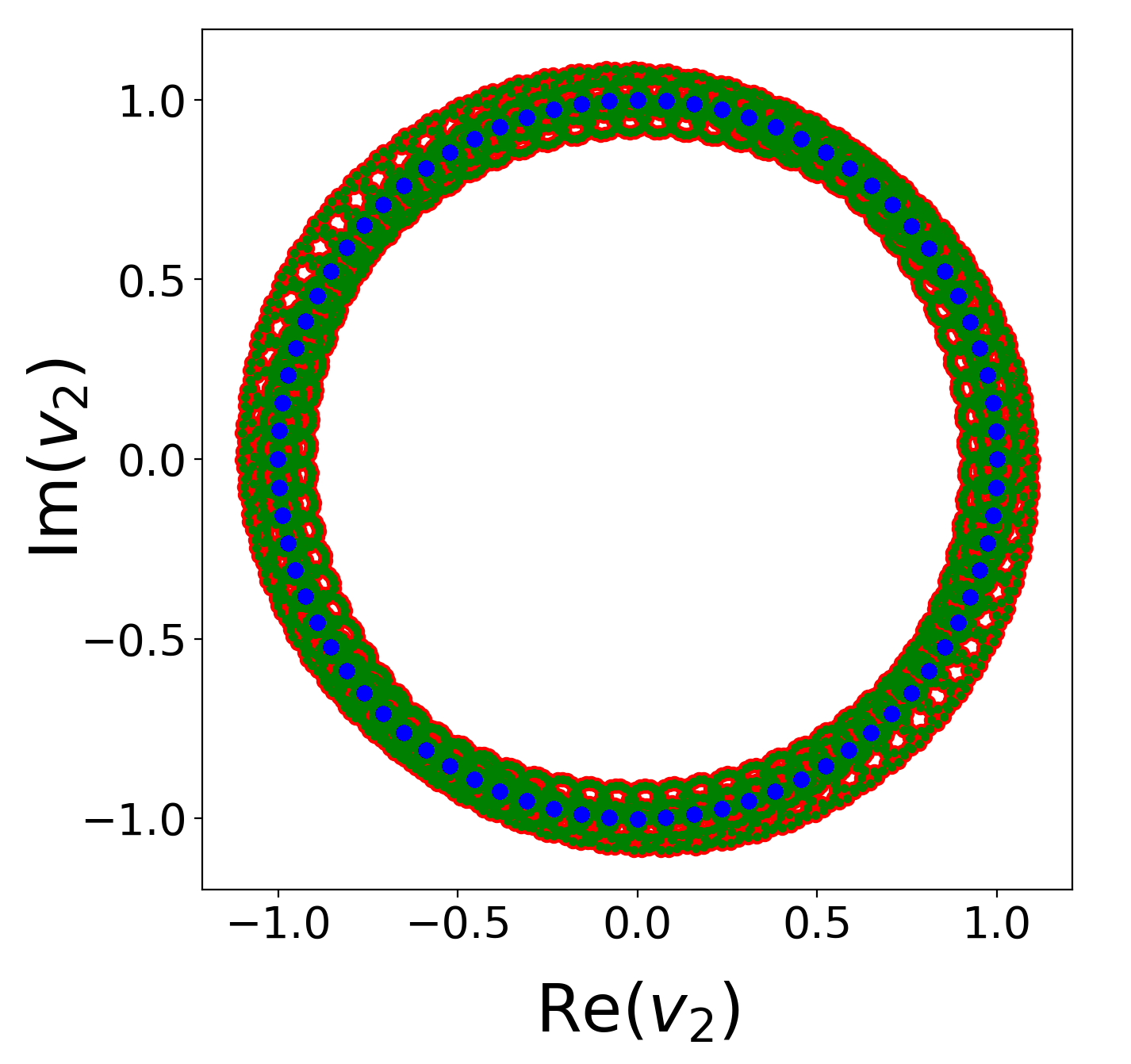}\vspace{-1.0em}
\par\end{raggedright}

\caption{phase space trajectories of $v_{2}^{(1)}$ calculated by Eq.(\ref{eq35})
in iteration 7 with maximum $n$ and $m$ between \{${-40,40}$\},
in clear agreement with Fig.9b. A comparison with Fig.3b shows how
the iteration reduces the fluctuation.}
\vspace{-2.5mm}\end{wrapfigure}%

\subsection{\label{sub:Results-of-solution}Results of solution for other initial
conditions}

When the iteration procedure is applied to initial coordinates for
$x_{0}=0$, and various \{$y_{0},p_{y0}$\}, the results from power
order $n_{s}=5$ (green) similar to Fig.11 are shown in Fig.12, showing
much better agreement with forward integration (red) than the contours
(magenta) from canonical perturbation theory \cite{Gustavson} calculated
at power order 8, even near $y_{0},p_{y0}=\{0,0.14\}$. 

\begin{figure}[b]
\vspace{-3.8mm}\includegraphics[bb=0bp 6bp 1330bp 290bp,clip,width=25.5cm]{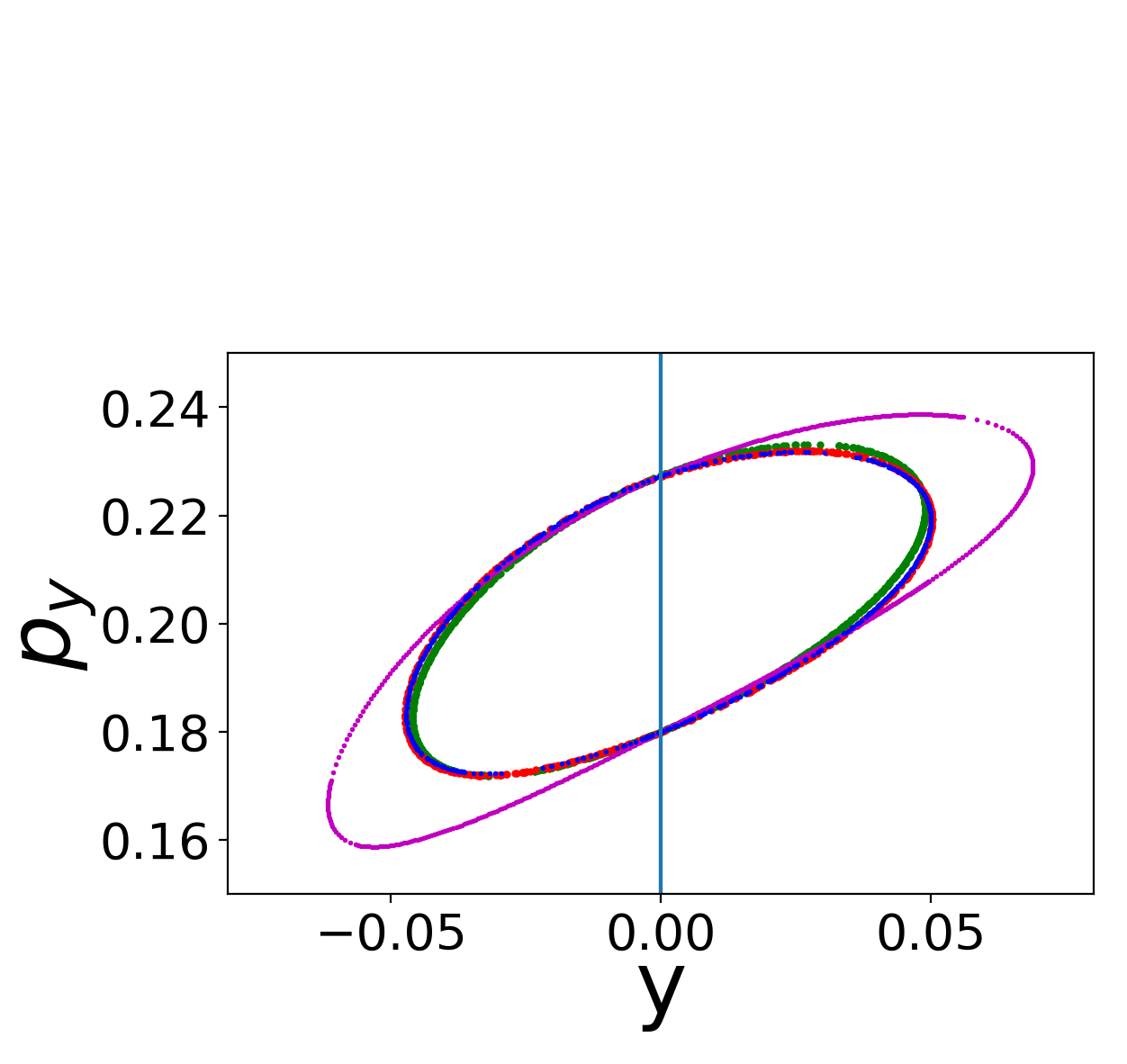}\vspace{-1.0em}

\caption{Poincare surface section $y,p_{y}$with $x=0$. Because energy conservation,
for energy E=0.0833, $p_{x}$ is determined from $x,y,p_{y}$, so
when the KAM invariant survives the trajectory passes through a surface
in the 3 dimension space of $x,y,p_{y}$. The trajectory crosses the
x=0 plane in one direction in the top half plane into the region $x>0$,
and later crosses the x=0 and comes back into the $x<0$ region. Thus
the trajectory crosses the plane by a curve shown as red points (the
forward integration $v_{1}^{(e)}$, $v_{2}^{(e)}$). The green dots
are the coordinates calculated from the inverse function of Eq.(\ref{eq:19})
with $v_{1}^{(1)}$, $v_{2}^{(1)}$ obtained in the last iteration
with power order $n_{s}=5$, corresponding to the red dots in Fig.8b,
in good agreement with the forward integration $v_{1}^{(e)}$, $v_{2}^{(e)}$.
The magenta dots are obtained from the well known canonical perturbation
theory based on Table IV in \cite{Gustavson}, calculated to 8th power
order. The result from square matrix clearly gives much better agreement
with forward integration. The blue dots are $\bar{v}_{01}$,$\bar{v}_{02}$,
 to be introduced in Section \ref{sec:Transform-Rotation-with}, in
nearly perfect agreement with forward integration (red).}
\end{figure}
Notice that at \{0,0.14\} the contour calculated from the canonical
perturbation theory at 8th power order jumps across the separatrix
and is marked as black contour circling around the fixed points near
\{0.25,0\} and(-0.3,0), while the numerical solution (red), the square
matrix solution (green) and the more accurate solution $\bar{v}_{02}$
(blue, to be introduced later in Section \ref{sec:Transform-Rotation-with})
remain at the same side, around the fixed points near \{0,0.2\} and
\{0,-0.2\}. 

In Fig.12, the square matrix calculation uses power order $n_{s}=5$
for all the initial positions. Except at $y_{0},p_{y0}=\{0,0.14\}$,
the error bar of $\bar{v}_{02}$ increases to 2.5\%, so we increases
power order to $n_{s}=7$ to reduce the error bar to about 1\%. When
we continue from \{0,0.14\} to \{0,0.135\} the iteration procedure
is no longe convergent for $n_{s}=7$ and $n_{v}=4$. 

When we keep $n_{s}=5$, we can continue from \{0,0.14\} down to as
low as \{0, 0.127\} but the error bar of $\bar{v}_{02}$ has increased
to 10\%. Numerical forward integration starts to show irregular behavior
at \{0, 0.123745\}, so the curve on the x=0 Poincare surface section
jumps back and forth between the two sides of the seperatrix. But
this value is sensitive to numerical integrator setup.

\section{\label{sec:Transform-Rotation-with}Transform Rotation with Fluctuation
into KAM invariant}

The procedure in Section \ref{sec:Iteration} results in the action-angle
variables $v_{1}^{(1)}$,$v_{2}^{(1)}$, which represent rigid rotations
with 4\% fluctuation. This fluctuation is related to the power order
of the square matrix $n_{s}=5$ and the number of left eigenvectors
used in the linear combination $n_{v}=4$, and the initial $y_{0},p_{y0}=0,0.18$.
However, the comparison with forward integration shows only 0.6\%
error. After the last iteration, $v_{1}^{(0)}$,$v_{2}^{(0)}$ represents
rigid rotations without fluctuation but it is not accurate solution.
$v_{1}^{(1)}$, $v_{2}^{(1)}$ represents more accurate solution but
with larger fluctuation. This allows us to derive a much more accurate
KAM invariant than $v_{1}^{(1)}$,$v_{2}^{(1)}$ using the relation
between $v_{1}^{(1)}$,$v_{2}^{(1)}$ and $v_{1}^{(0)}$,$v_{2}^{(0)}$.

The relation is derived as follows. Substitute the expression of zeroth
order action Eq.(\ref{eq:22}) into the first order action Eq.(\ref{eq35}),
use the normalized actio-angle variables $\bar{v}_{1},\bar{v}_{2}$,
substitute $e^{i\theta_{1}^{(0)}(t)},e^{i\theta_{2}^{(0)}(t)}$ by
$\bar{v}_{1}^{(0)},\bar{v}_{2}^{(0)}$ , and use the iteration result
of $v_{1}^{(1)}\approx v_{1}^{(e)}$, $v_{2}^{(1)}\approx v_{2}^{(e)}$
as shown at the end of Section \ref{sub:Numerical-Examples} about
Fig.8, we get
\begin{equation}
\begin{split} & \bar{v}_{1}^{(e)}\approx\bar{v}_{1}^{(1)}=\bar{v}_{1}^{(0)}e^{i\thinspace\sum_{n,m}\widetilde{\theta}_{1nm}(\bar{v}_{1}^{(0)})^{n}(\bar{v}_{2}^{(0)})^{m}}\\
 & \bar{v}_{2}^{(e)}\approx\bar{v}_{2}^{(1)}=\bar{v}_{2}^{(0)}e^{i\thinspace\sum_{n,m}\widetilde{\theta}_{2nm}(\bar{v}_{1}^{(0)})^{n}(\bar{v}_{2}^{(0)})^{m}}
\end{split}
\label{eq:36}
\end{equation}

Eq.(\ref{eq:36}) expresses the exact action in terms of the zeroth
order action. Given the zeroth order function $\bar{v}_{1}^{(0)},\bar{v}_{2}^{(0)}$
as rigid rotations, Eq.(\ref{eq:36}) gives $\bar{v}_{1}^{(1)},\bar{v}_{2}^{(1)}$
as perturbed rigid rotations with fluctuation, which more closely
represents the motion $\bar{v}_{1}^{(e)},\bar{v}_{2}^{(e)}$ .

Notice the first approximate equal signs in the left hand side of
Eq.(\ref{eq:36}) are valid only if $\bar{v}_{1}^{(0)},\bar{v}_{2}^{(0)}$
are on unit circle, we can use the inverse function of Eq.(\ref{eq:36})
to test if $\bar{v}_{1}^{(0)}(t),\bar{v}_{2}^{(0)}(t)$ do represent
rigid rotations. In other words, the inverse function $\bar{v}_{1}^{(0)},\bar{v}_{2}^{(0)}$
of Eq.(\ref{eq:36}), expressed as a function of $\bar{v}_{1}^{(e)},\bar{v}_{2}^{(e)}$,
as polynomials of $x,p_{x},y,p_{y}$, should be a much more accurate
action approximation than $\bar{v}_{1}^{(1)},\bar{v}_{2}^{(1)}$.
In addition, when we are searching for KAM invariants we always assume
$x,p_{x},y,p_{y}$ are on the exact trajectory, so in the following
we replace the notation $\bar{v}_{1}^{(e)},\bar{v}_{2}^{(e)}$ by
$\bar{v}_{1},\bar{v}_{2}$.

To calculate this inverse function, we first neglect the small terms
in the right hand side of Eq.(\ref{eq:36}) so we have $\bar{v}_{1}^{(0)}\approx\bar{v}_{1}^{(1)}\approx\bar{v}_{1}^{(e)}=\bar{v}_{1}$,
$\bar{v}_{2}^{(0)}\approx\bar{v}_{2}^{(1)}\approx\bar{v}_{2}^{(e)}=\bar{v}_{2}$.
Replace $\bar{v}_{1}^{(0)},\bar{v}_{2}^{(0)}$ by $\bar{v}_{1},\bar{v}_{2}$
in the exponents in Eq.(\ref{eq:36}), we found
\begin{equation}
\begin{split} & \bar{v}_{01}=\bar{v}_{1}^{(0)}=\bar{v}_{1}^{(e)}e^{-i\thinspace\sum_{n,m}\widetilde{\theta}_{1nm}(\bar{v}_{1}^{(0)})^{n}(\bar{v}_{2}^{(0)})^{m}}\approx\\
 & \bar{v}_{1}e^{-i\thinspace\sum_{n,m}\widetilde{\theta}_{1nm}(\bar{v}_{1})^{n}(\bar{v}_{2})^{m}}\approx\bar{v}_{1}\thinspace(1-i\thinspace\sum_{n,m}\widetilde{\theta}_{1nm}(\bar{v}_{1})^{n}(\bar{v}_{2})^{m})\\
 & \bar{v}_{02}=\bar{v}_{2}^{(0)}=\bar{v}_{2}^{(e)}e^{-i\thinspace\sum_{n,m}\widetilde{\theta}_{2nm}(\bar{v}_{1}^{(0)})^{n}(\bar{v}_{2}^{(0)})^{m}}\approx\\
 & \bar{v}_{2}e^{-i\thinspace\sum_{n,m}\widetilde{\theta}_{2nm}(\bar{v}_{1})^{n}(\bar{v}_{2})^{m}}\approx\bar{v}_{2}\thinspace(1-i\thinspace\sum_{n,m}\widetilde{\theta}_{2nm}(\bar{v}_{1})^{n}(\bar{v}_{2})^{m})
\end{split}
\label{eq:37}
\end{equation}
\begin{figure}
\vspace{-1.0em}\includegraphics[width=1\columnwidth]{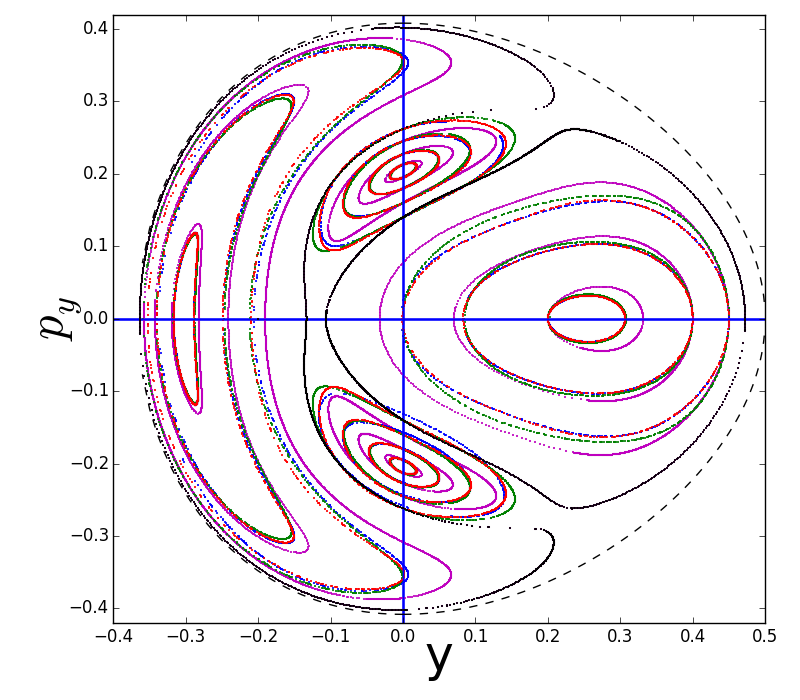}\vspace{-1.0em}\caption{Poincare surface section $x=0$ for trajectory with initial condition
$x_{0}=0$, and $y_{0},p_{y0}=$ \{0, 0.195\}, \{0,0.18\}, \{0,0.16\},
\{0,0.14\}, \{0.2,0\}, \{0.4,0\}, \{0.45,0\}. Red: forward numerical
integration. Green: constant contours of $v_{1},v_{2}$ given by Eq.(\ref{eq:19}).
Blue: constant contours of $\bar{v}_{01},\bar{v}_{02}$ given by Eq.(\ref{eq:37})
(see Section \ref{sec:Transform-Rotation-with}). Magenta: contours
from Table IV of \cite{Gustavson}. Notice at \{0,0.14\} the contour
calculated from the canonical perturbation theory at 8th power order
jumps across the separatrix and is marked as black contour. The dashed
black line is the energy limit $\frac{1}{2}p_{x}^{2}+\frac{1}{2}y^{2}-\frac{1}{3}y^{3}=E$
for $E=0.0833$. The solution for $y_{0},p_{y0}=$ \{0.2,0\}, \{0.4,0\}
and \{0,0.195\} are found in the same way as \{0,0.18\}. The solutions
for \{0,0.16\} can also be found the sam way, but here it is found
by starting the iteration with the linear combinations obtained from
\{0,0.18\} case. The dominance of the $\omega_{2}$ line in $v_{2}^{(1)}$
is established from the start so the convergence process starts immediately.
The 0.14 case follows the solution for 0.16 by first going to 0.145,
then 0.143. The solution for \{0.45,0\} is found by starting iteration
from the solution for \{0.4,0\}. We see that the blue curve (see Section
\ref{sec:Transform-Rotation-with}) starts to show a small deviation
from the numerical integration near $y_{0},p_{y0}=$ \{0,-0.14\} but
not at \{0,0.14\}. This asymmetry is due to the fact the solution
process is not symmetric for upper and lower half plane: we start
the initial point at \{0, 0.14\}, not \{0,-0.14\} so the errors are
different between upper and lower half plane. }

\vspace{0.0em}
\end{figure}
Here we denote the $\bar{v}_{1}^{(0)},\bar{v}_{2}^{(0)}$ using a
different notation $\bar{v}_{01},\bar{v}_{02}$ to indicate the distinction
between the variable $\bar{v}_{1}^{(0)},\bar{v}_{2}^{(0)}$ in Eq.(\ref{eq:36})
and $\bar{v}_{01},\bar{v}_{02}$ in Eq.(\ref{eq:37}). In Eq.(\ref{eq:36})
we consider them as a simple expression as in Eq.(\ref{eq:22}) representing
rigid rotations in the phase space of $\theta_{1}^{(0)},\theta_{2}^{(0)}$.
Then Eq.(\ref{eq:36}) gives a more accurate representation of the
motion, as two independent perturbed rigid rotations. In Eq.(\ref{eq:36})
$\bar{v}_{1}^{(0)},\bar{v}_{2}^{(0)}$ are calculated from Eq.(\ref{eq:22})
as function of $\theta_{1}^{(0)},\theta_{2}^{(0)}$, or $x^{(0)},p_{x}^{(0)},y^{(0)},p_{y}^{(0)}$
, while $\bar{v}_{1}^{(1)},\bar{v}_{2}^{(1)}$ are function of $x^{(1)},p_{x}^{(1)},y^{(1)},p_{y}^{(1)}$.
Eq.(\ref{eq:36}) is used to find solution without forward integration.

As a comparison, in Eq.(\ref{eq:37}), $\bar{v}_{01},\bar{v}_{02}$
are more complicated function of $x,p_{x},y,p_{y}$ than $\bar{v}_{1},\bar{v}_{2}$
, they represent the much more accurate approximation to rigid rotations.
In Eq.(\ref{eq:37}) $\bar{v}_{1},\bar{v}_{2}$ are calculated from
Eq.(\ref{eq:19}) as function of $x^{(e)},p_{x}^{(e)},y^{(e)},p_{y}^{(e)}=x,p_{x},y,p_{y}$.
Eq.(\ref{eq:37}) gives a KAM invariant for a known solution.

In Fig.13a we show the phase space of $\bar{v}_{02}$ as given by
Eq.(\ref{eq:37}). Clearly this represents a rigid rotation with much
less fluctuation than $\bar{v}_{2}^{(1)}$ as shown in Fig.9b. We
remark here that clearly the solution $\bar{v}_{01},\bar{v}_{02}$
in Eq.(\ref{eq:37}) is not a Taylor expansion, its exponent is a
function with both positive and negative powers of the polynomials
$\bar{v}_{1},\bar{v}_{2}$, i.e., a Laurent series of $\bar{v}_{1},\bar{v}_{2}$.
When we carry out Taylor expansion of $\bar{v}_{02}$ to $n_{s}=5$
order, the result has a huge fluctuation. Our numerical test shows
only if we expand up to power order of 17 we are able to keep the
fluctuation to 0.6\% as shown in Fig.13a. From this, we expect the
solution should be in the general form Eq.(\ref{eq:37}), where the
power order $n_{s}=5$ of $\bar{v}_{1},\bar{v}_{2}$ is not required
to be very high to reach high precision. It is not appropriate to
use power expansion for the KAM invariant $\bar{v}_{01},\bar{v}_{02}$
to achieve high precision.

To compare the errors of several action-angle variables we derived,
in Fig.13b we plot the details of the contours of the x=0 Poincare
surface section for initial $y_{0},p_{y0}=0,0.18$ in Fig.12. It is
very clear from this plot, $\bar{v}_{2}^{(1)}$ (green) the first
order approximation is much more close to the forward integration
(red) than the magenta curve derived from the canonical perturbation
theory. The more accurate $\bar{v}_{02}$ (blue) given by Eq.(\ref{eq:37})
is even much more close to the red than the green. This demonstrates
the high precision of KAM invariant Eq.(\ref{eq:37}).
\begin{figure}[t]
\vspace{-1.0em}\includegraphics[width=4.41cm]{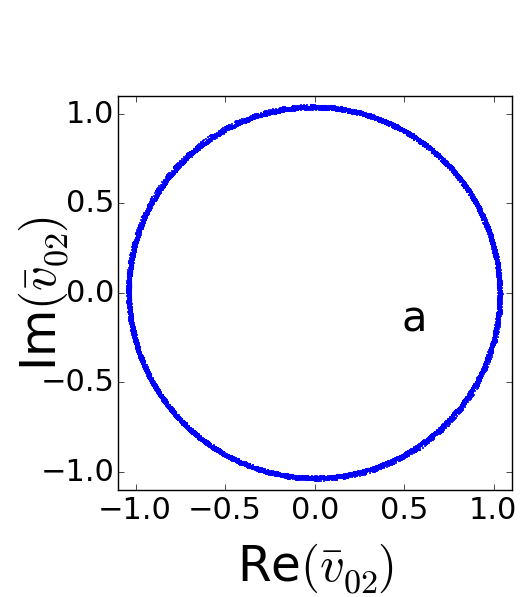}\includegraphics[width=4.23cm]{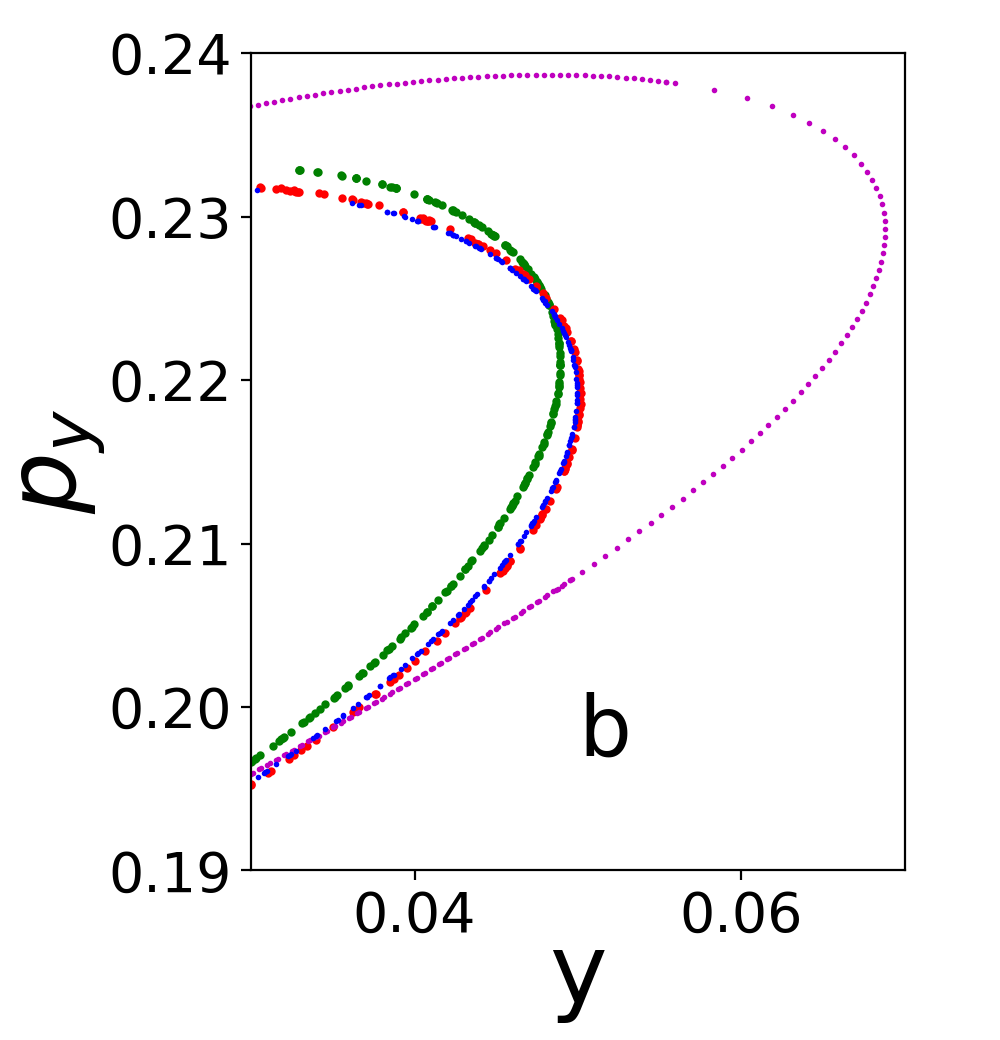}\vspace{-1.0em}\caption{For the trajectory with initial condition $x_{0}=0$, and $y_{0},p_{y0}=$
\{0,0.18\}, power order $n_{s}=5$. a. phase space of $\bar{v}_{02}$
given by Eq.(\ref{eq:37}) with $x,p_{x},y,p_{y}$ in $\bar{v}_{1},\bar{v}_{2}$
calculated from forward integration, nearly a perfect circle.\qquad{}b.
A detailed magnified part in Fig.12 for error comparison of different
KAM invariant approximations on Poincare surface section $x=0$. Red:
forward numerical integration. Green: constant contours of $v_{1},v_{2}$
given by Eq.(\ref{eq:19}). Blue: constant contours of $\bar{v}_{02}$
given by Eq.(\ref{eq:37}). Magenta: contours from Table IV of \cite{Gustavson}
based on canonical perturbation theory. Clearly $\bar{v}_{02}$ is
much more precise than previous theory. }
\end{figure}

\section{Conclusion}

We developed a perturbation theory for nonlinear dynamics on resonances
based on using the liner combinations of left eigenvectors in the
degenerate Jordan chains of a square matrix as the zeroth order approximate
action-angle variables to find highly accurate approximation of the
KAM invariants. The solution is not in the form of a power series,
but in the form of an exponential function with rational function
as its exponent, more like a Laurent series rather than a Taylor series.
To achieve high precision, the required power order for the action-angle
variables in the exponent is much less stringent than the required
power order for the power series expansion to achieve the same precision. 

The solution is found by an iteration procedure. In each iteration
step, we need to solve a set of linear equations to improve the accuracy.
Numerical study shows, for the example of Henon-Heiles problem, the
iteration converges in much of the region where the KAM invariants
persist. While the result of the canonical perturbation theory in
the example still gives contours in the region where the orbits fall
into chaotic behavior (see Fig.10 of \cite{Gustavson} and the black
curve in Fig.12 of Section \ref{sub:Results-of-solution}), the iteration
based on the square matrix method is no longer convergent when approaching
the stability boundary. This gives a way to get information about
the stability boundary, and also raised the question about the relation
of the convergence region and the stability boundary. Whether there
is an analytical answer on this question remains to be a very interesting
and important issue.

While for the canonical perturbation theory the measure of the perturbation
is given by the amplitude of the nonlinear terms, for the perturbation
theory developed for square matrix the measure of the perturbation
is given by the fluctuation relative to the amplitude of the action-angle
variables in the initial trial iteration. Hence when approaching the
stability boundary the reduction of the step size can reduce the fluctuation
and may lead to convergence and improve the precision with increased
power order of $n_{s}$, or increased number of left eigenvectors
$n_{v}$. Hence the convergence and precision seem to be determined
by the ratio of the increase of the fluctuation over the step size,
instead of being determined by the amplitude of the perturbation only.
In this sense the meaning of perturbation here is different from the
conventional meaning, or may be even beyond perturbation, hence may
have the potential to explore the area with increased amplitude or
perturbation.
\begin{acknowledgments}
The author would like to thank Prof. C.N. Yang for discussion and
encouragement. Thank Dr. G. Stupakov for his many comments, suggestions
and discussion on this paper. We also would like to thank Prof. Yue
Hao for discussion and comments on the manuscript, and for providing
TPSA programs to construct the square matrixes. This was funded by
DOE under Contract No. DE-SC0012704.\end{acknowledgments}

\end{document}